\documentclass[sigconf,nonacm]{acmart}
\AtBeginDocument{%
  }

\setcopyright{rightsretained}
\copyrightyear{2024}
\acmYear{2024}

\usepackage{bbding}
\usepackage{threeparttable}

\newcommand{\sysname}{\texttt{Vul-BinLLM~}}

\begin{document}

\title{VulBinLLM: LLM-powered Vulnerability Detection for Stripped Binaries}


\author{Nasir Hussain}
\affiliation{%
  \institution{University of California, Los Angeles}
  \country{}
}
\email{nasirhm@ucla.edu}

\author{Haohan Chen}
\affiliation{%
  \institution{University of California, Los Angeles}
  \country{}
}
\email{heyohan@ucla.edu}

\author{Chanh Tran}
\affiliation{%
  \institution{University of California, Los Angeles}
  \country{}
}
\email{ctran0014@ucla.edu}

\author{Philip Huang}
\affiliation{%
  \institution{University of California, Los Angeles}
  \country{}
}
\email{philiph930@g.ucla.edu}

\author{Zhuohao Li}
\affiliation{%
  \institution{University of California, Los Angeles}
  \country{}
}
\email{zhuohaol@g.ucla.edu}

\author{Pravir Chugh}
\affiliation{%
  \institution{University of California, Los Angeles}
  \country{}
}
\email{pravirchugh@ucla.edu}

\author{William Chen}
\affiliation{%
  \institution{University of California, Los Angeles}
  \country{}
}
\email{billchen314@g.ucla.edu}

\author{Ashish Kundu}
\affiliation{%
  \institution{Cisco Research}
  \country{}
}
\email{ashkundu@cisco.com}

\author{Yuan Tian}
\affiliation{%
  \institution{University of California, Los Angeles}
  \country{}
}
\email{yuant@ucla.edu}

\renewcommand{\shortauthors}{Hussain et al.}

\begin{abstract}
Recognizing vulnerabilities in stripped binary files presents a significant challenge in software security. Although some progress has been made in generating human-readable information from decompiled binary files with Large Language Models (LLMs), effectively and scalably detecting vulnerabilities within these binary files is still an open problem. This paper explores the novel application of LLMs to detect vulnerabilities within these binary files. We demonstrate the feasibility of identifying vulnerable programs through a combined approach of decompilation optimization to make the vulnerabilities more prominent and long-term memory for a larger context window, achieving state-of-the-art performance in binary vulnerability analysis. Our findings highlight the potential for LLMs to overcome the limitations of traditional analysis methods and advance the field of binary vulnerability detection, paving the way for more secure software systems.

In this paper, we present \sysname, an LLM-based framework for binary vulnerability detection that mirrors traditional binary analysis workflows with fine-grained optimizations in decompilation and vulnerability reasoning with an extended context. In the decompilation phase, \sysname adds vulnerability and weakness comments without altering the code structure or functionality, providing more contextual information for vulnerability reasoning later. Then for vulnerability reasoning, \sysname combines in-context learning and chain-of-thought prompting along with a memory management agent to enhance accuracy. Our evaluations encompass the commonly used synthetic dataset Juliet to evaluate the potential feasibility for analysis and vulnerability detection in C/C++ binaries. Our evaluations show that \sysname is highly effective in detecting vulnerabilities on the compiled Juliet dataset.
\end{abstract}

\maketitle
\pagestyle{plain}

\section{Introduction}
\label{intro}
Binary analysis is pivotal for program analysis and provides a deep understanding of how executable files operate. In many cases, the source code is not available for analysis \cite{shoshitaishvili2016sok, pang2021sok}, especially with proprietary or legacy software, which is common in commercial software (e.g. Microsoft Windows, Adobe Acrobat), firmware in IoT devices, and third-party libraries \cite{zhao2023uvscan} where vendors often withhold source code to protect intellectual property. Binary analysis allows security engineers to directly analyze the compiled binary, enabling them to understand how the program functions and identify potential vulnerabilities and weaknesses.
Traditional binary analysis tools, such as Ghidra~\cite{ghidra}  IDA Pro~\cite{idapro}, and Angr~\cite{wang2017angr} analyze binary code by translating it into assembly language and providing higher-level representations to facilitate understanding. This is accomplished through two main processes: disassembly and decompilation. Disassembly converts machine code into assembly language, while decompilation translates this code into a high-level programming language. However, this process inherently leads to information loss, for reasons such as loss of high-level constructs, compiler optimizations, and symbol and debug information removal. 
Thus, this leads to lots of ambiguity in the decompiled code. 
Additionally, learning and mastering reverse engineering require significant effort and expertise. The raw binary format and the complexity of the analysis process make it challenging and time-consuming for security analysts to detect vulnerabilities in the binaries. Therefore, automating the analysis process is essential to efficiently detect vulnerabilities in binaries, reduce manual efforts, and improve the speed of vulnerability identification.

Recent developments in large language models (LLMs) \cite{openai2024chatgpt, touvron2023llama, codellama, anthropic2024claude, codestral} present a promising avenue for addressing the challenges of binary analysis. LLMs have demonstrated capabilities in code summarization and generation \cite{nam2024using, madaan2024self, scheurer2023training, kang2023large, neelakantan2022text, muennighoff2023octopack, xia2023automated, steenhoek2023empirical, du2024vul}. LLM agents~\cite{talebirad2023multi}, which integrate LLMs as a component in the workflow, enable interactions that closely mimic expert-level intelligence. This emergence and rapid adoption of LLMs in code analysis and code copilots raises a critical research question: \textit{Can large language models, with fine-grained prompt engineering and specialized optimizations in system design, assist security and software engineers to effectively reason about vulnerabilities in binary code?}

\textit{Objective of this work:} We aim to provide an LLM-powered approach to enable efficient, and scalable vulnerability analysis for the binary code. However, applying LLMs within binary analysis retains several challenges. First, compilation and decompilation causes a loss of contextual information for the code, which makes it challenging to identify vulnerabilities on the binary level. While source code vulnerabilities are typically classified using standards like Common Vulnerabilities and Exposures (CVEs) \cite{cve} and Common Weakness Enumeration (CWEs) \cite{cwe}, translating these definitions to the binary level is challenging. Many reported CVEs involve complex projects with custom-built toolchains. These projects include build configurations and compilation settings that optimize the binary for various use cases, resulting in an even higher loss of information during the compilation phase. Thus, it is difficult to map the binary representation to its source code. Complex compilation processes add a layer of difficulty in accurately tracking and detecting vulnerable code within binary files. 
Second, binary code is much more abstract than natural language. Since LLMs rely on probabilities to predict in an auto-regressive manner, they often have difficulty  extracting meaningful information from low-level representations directly \cite{tan2024llm4decompile}. Reverse engineering tools are able to help extract useful information, but the output usually lacks semantic information like variable names and comments. Third, LLMs might suffer from hallucinations, especially with longer input. This issue is orders of magnitude worse in binary analysis, where the lack of context and semantic clues makes it harder for LLMs to interpret the actual behavior and vulnerabilities of the code reliably. Finally, binary files are usually very large, posing challenges for the limited context window of LLMs. As a result, as far as we know, state-of-the-art LLM-assisted binary analysis solutions are usually evaluated with small synthetic datasets, and cannot handle real-world binaries. 

To address these challenges, we propose \sysname, an end-to-end LLM-powered binary vulnerability analyzer. As far as we know, \sysname is a LLM-assisted binary vulnerability analysis tool that detects vulnerabilities from compiled binaries. We achieve this goal by effectively allowing the LLM's to analyze a binary file that far exceeds its context window and attaching a function analysis queue that reduces hallucinations in the memory management agent to ensure complete analysis of the binary file. 


\sysname employs a structured workflow to assess and optimize the binary analysis process iteratively, emulating the approach of a security expert. It features an optimized decompiler that enhances decompiled code by appending vulnerability-specific comments, as well as prompt engineering to enhance vulnerability reasoning. By decomposing the traditional binary analysis workflow, \sysname integrates insights from the extensive decompiled source code, offering an effective memory management approach to improve LLM analysis. In order to reduce hallucinations from processing of the code within the limited context window, we attach a function queue with the LLM's extended memory. 

We evaluate \sysname on binaries on commonly used synthetic vulnerability datasets i.e Juliet. \sysname is able to beat the state-of-art tools and frameworks for stripped synthetic data binaries. VulBinLLM can detect CWEs for Juliet with appropriate information to justify the presence of such vulnerability in code.

We summarize our contributions for building a LLM-Powered binary analysis framework as follows: 
\vspace{-\topsep} 
\begin{itemize}
    \item We build an LLM-assisted binary vulnerability analysis tool tailored for CWE detection on decompiled files by integrating a memory management system and a function analysis queue, enabling it to analyze complex binaries.
    \item With our analysis, we provide insights into how LLMs can analyze vulnerabilities on the binary level without much contextual knowledge and be able to reconstruct them for vulnerability detection.
    \item With our evaluations, we achieved approximately ~10\% increased accuracy in detecting stripped synthetic code vulnerabilities.
\end{itemize}

\section{Background}
\label{background}

\begin{figure*}[!t]
\centering
\includegraphics[width=0.75\linewidth]{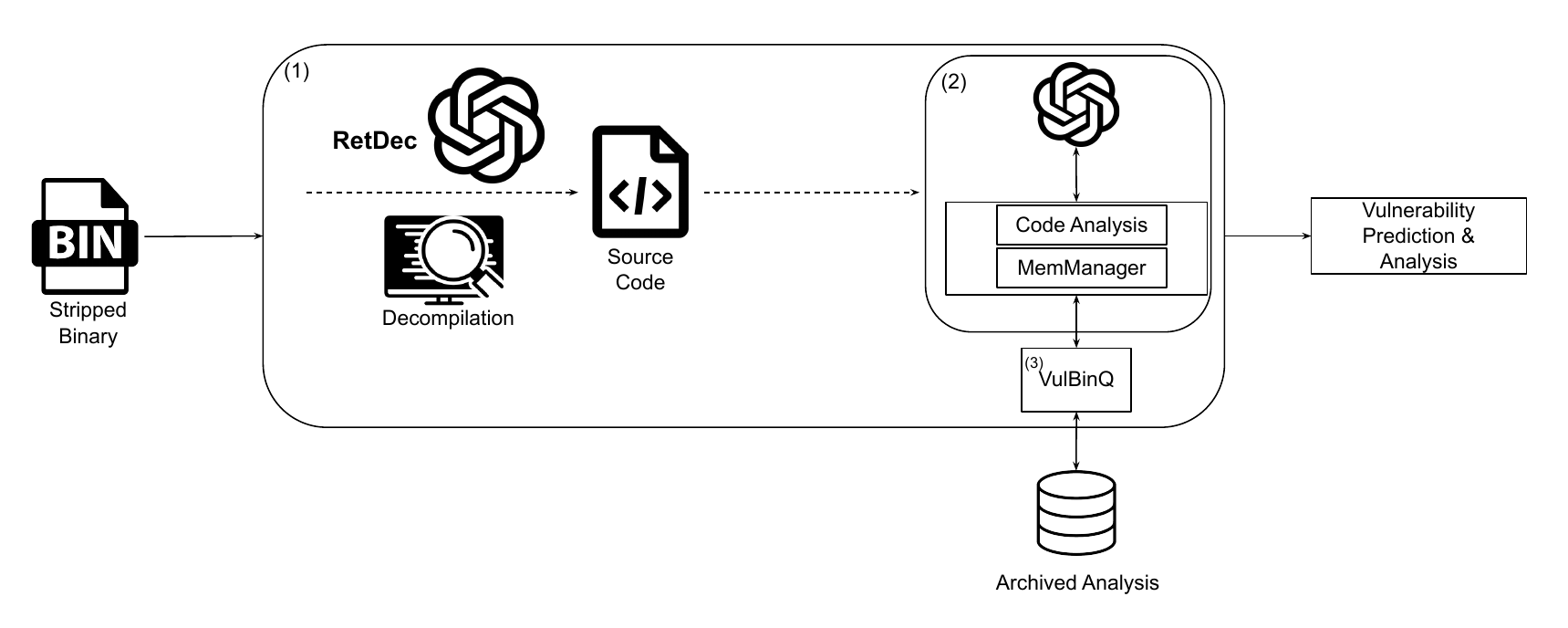}
\caption{Workflow of \sysname: (1) Binary files are decompiled into source code, where an LLM-assisted decompiler enriches the code with contextual information for vulnerability detection. (2) The decompiled source code is then analyzed by \sysname for vulnerability analysis which has an archival storage to store analysis, The analyzer then provides a comprehensive vulnerability detection result (3) VulBinQ: It features an additional queue that manages the functions that are to be analyzed using \sysname and serves as middleware between the Archived Analysis and \sysname.}
\label{fig_sim}
\end{figure*}

\subsection{Reserve Engineering for Vulnerability Analysis}

Reverse engineering plays a critical role in vulnerability analysis, malware detection, and overall system security. In system security, reverse engineering refers to the process by which an analyst examines a binary executable to recover the design and implementation details necessary for understanding the program's functionality. This process can be applied in various contexts, such as malware analysis, vulnerability discovery, or firmware analysis, with the specific output varying according to the context. However, the core goal remains the same: reconstructing the program logic and identifying the conditions required to reach specific code locations, which could reveal bugs or suspicious behaviors, especially in the case of malicious binaries.

While automation has significantly advanced in areas like host-based and network-based attack detection, malware classification, and phishing detection, binary reverse engineering is still primarily driven by highly skilled human experts. Although tools for unpacking, disassembly, emulation, and binary similarity comparison assist in the process, the task of fully understanding the code remains a predominantly human effort. This manual work demands a deep level of expertise and is both time-consuming and labor-intensive. Unfortunately, the shortage of expert reverse engineers is a bottleneck, especially considering the increasing volume of software that requires analysis.

Initiatives like the DARPA Cyber Grand Challenge (CGC) have pushed forward the development of autonomous systems capable of analyzing binaries and identifying vulnerabilities. Despite these advancements, current automated solutions still struggle to match human-expert level depth of reasoning and intuition in reverse engineering tasks. As such, reverse engineering continues to be a crucial component of software vulnerability discovery, relying heavily on a combination of human skill and machine assistance.

\subsection{Binary Analysis and Program Analysis}
\label{binary_program_analysis}

Binary analysis can be broadly categorized into dynamic and static binary analysis. Dynamic analysis examines program behavior during execution, while static analysis involves examining binary code without executing it. In this paper, unless otherwise noted, we focus exclusively on static binary analysis.

Traditional static binary analysis begins with binary acquisition and format inspection, followed by systematic layers of examination, including disassembly, control flow analysis, and data flow analysis. Disassembly and decompilation converts binary code into human-readable assembly instructions or source code. Control flow analysis maps the potential execution paths by identifying decision points and jumps. Data flow analysis tracks data movement and transformations to uncover vulnerabilities such as buffer overflows or memory leaks. Additional techniques like call graph analysis, help visualize relationships between functions and identify entry and exit points. Symbolic execution explores all possible execution paths by using symbolic rather than concrete values, uncovering potential vulnerabilities missed by traditional methods. Static analysis tools also commonly incorporate pattern matching to detect known vulnerabilities, which can accelerate the analysis process. However, this comprehensive analysis usually requires significant expertise and extensive training.

\begin{figure}[h]
\centering
\includegraphics[width=\linewidth]{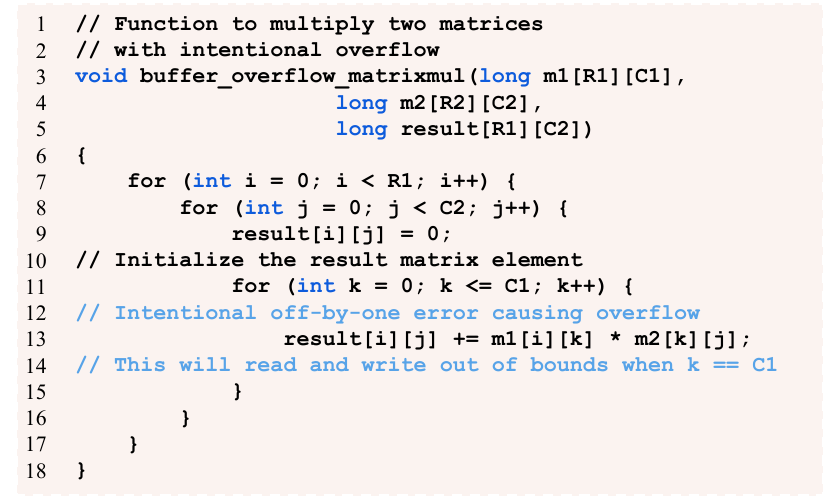}
\caption{An example of a buffer overflow vulnerability: 
the detection of this vulnerability relies on human expertise in security review. Human experts are able to detect the above vulnerability, but the subtley of the vulnerability may lead to difficulty in detection by an LLM.}
\label{fig_buff}
\end{figure}

\subsection{CVE and CWE}
Existing vulnerability classification systems, such as Common Vulnerabilities and Exposures (CVE) \cite{cve} and Common Weakness Enumeration (CWE) \cite{cwe}, provide standardized frameworks for categorizing and managing security issues. CVE is a system used to describe and report specific vulnerabilities in real-world applications (e.g. Google Chrome, Android), with each reported vulnerability assigned a unique identifier (CVE ID). A single CVE ID may correspond to multiple distinct code snippets representing different instances of the same vulnerability. CWE serves as a classification system that organizes common software and hardware security weaknesses. Each weakness type is assigned a unique identifier (CWE ID), providing a broad taxonomy of security flaws. Code behaviors within the same CWE category can vary significantly. For instance, CWE-416 (Use After Free) \cite{cwe416} denotes improper use of memory after it has been freed, which may result from issues such as race condition mismanagement (e.g., CVE-2023-30772 \cite{cve-2023-30772}) or incorrect reference counting, leading to premature object destruction (e.g., CVE-2023-3609 \cite{cve-2023-3609}). While CWE offers a higher-level understanding of weaknesses, CVE provides specific instances of vulnerabilities in real-world systems. CWE and CVE are widely used in the standardized approach for categorizing and managing security issues.

\begin{table}[ht]
\centering
\begin{tabular*}{\columnwidth}{@{\extracolsep{\fill}} l c c} 
 \toprule
 Datasets & \# of CWEs &  \# of Test Cases \\ 
 \midrule
 Juliet C/C++ \cite{juliet_c_cpp} & 118 & 90,000+ \\
 SARD \cite{sard} & 200+  & 100,000+ \\
 Big-Vul \cite{big_vul}& 10+ critical CWEs & 6,800 \\
 Devign \cite{devign}& 5+ memory-related CWEs & 7,000+ \\
 REVEAL \cite{reveal}& 10+ memory safety CWEs & 3,000+ \\
 MVD \cite{mvd}& 10+ critical CWEs & 1,000+ \\
 Vul4J \cite{vul4j}& 10+ Java-specific CWEs & 500+ \\
 \bottomrule
\end{tabular*}
\caption{Overview of widely-used CWE datasets in vulnerability detection. Juliet C/C++ and SARD each contain over 100 CWEs  for in-depth testing in C/C++ and Java. NVD includes over 150 CWEs linked to public CVEs. Big-Vul offers critical vulnerabilities from open-source projects. Devign and REVEAL focus on memory-related weaknesses in C/C++. MVD and Vul4J cater to critical and Java-specific vulnerabilities.}
\label{tab_cwe}
\end{table}

\section{Method}
\label{method}

\subsection{Overview}

\sysname breaks the problem down into two steps: 1) the reconstruction of information optimized for vulnerability detection from binary files, 2) the analysis of decompiled code to infer vulnerabilities in terms of CWEs. In particular, the second step poses a unique challenge when the decompiled code exceeds the context window for the LLM, leading to an inability for vulnerability analysis of a binary file. To further enhance the problem statement we present the following Research Questions (RQ) to analyze:  
\textbf{RQ1:} Does LLM-powered human-readable code restoration for functionality description allow for vulnerability detection using LLMs? 

\textbf{RQ2:} Can we optimize the restoration to make the vulnerability features more prominent such that the LLM's ability to detect vulnerabilities is improved?

To address these RQs, we propose \sysname, a LLM-powered binary analysis framework. \sysname is the first framework that focuses on recovering syntactic information to highlight vulnerable features using LLMs. 

In this section, we present the methodologies behind \sysname. It employs a structured workflow to first generate syntactic information from the binary code, making the vulnerable features more prominent within the source code, to allow the LLM to detect vulnerabilities using this syntactic information. Then, it iteratively assesses the source code with an extended memory structure to analyze each function in accordance to the control flow, emulating the approach of a security expert. By decomposing the traditional binary analysis workflow, \sysname integrates insights from the decompiled source code by embedding appropriate information during the decompilation stage and then utilizing it for vulnerability detection, offering an integrated approach to improve vulnerability analysis. Detailed discussions are provided in Section \ref{method1}.

First, we summarize the challenges of applying LLMs to binary analysis and describe how \sysname addresses them:
\begin{itemize}
    \item \textbf{Feature Definition Code Generation: } Binary code, especially in stripped binaries, lacks the high-level abstractions and semantic information present in source code. This makes it difficult for LLMs, which are primarily trained on text and source code, to understand the underlying logic and identify potential vulnerabilities.
    
    \item \textbf{Data Scarcity: } Training effective LLMs for binary vulnerability detection requires large datasets of labeled binary code with known vulnerabilities. However, such datasets are relatively scarce compared to source code datasets, which can hinder the development and evaluation of LLM-based approaches.

    \item \textbf{Limited Contextual Awareness: } LLMs may struggle to understand the broader context in which a particular code segment operates. This can lead to inaccurate vulnerability assessments, as the model may not fully grasp the implications of a specific code pattern within the larger program.
\end{itemize}

\begin{table*}[ht]
\centering
\begin{tabular*}{\textwidth}{@{\extracolsep{\fill}} l  l  l  l  l  l  l l} 
 \toprule
  Models & Creator & \# Parameters & Modality & Max Tokens & Function & Vul-1 \footnotemark & Vul-2\\ 
\midrule
  GPT-4 \cite{gpt4}  & OpenAI & 1.7T & text & 32K & \Checkmark & \Checkmark & \XSolidBrush\\ 
  \hline
  ChatGPT \cite{openai2024chatgpt}& OpenAI & 175B & text & 16K & \Checkmark & \Checkmark & \XSolidBrush\\ 
  \hline
  Claude 3.5 \cite{anthropic2024claude} & Anthropic & 175B & text & 32K & \Checkmark & \Checkmark & \XSolidBrush\\ 
  \hline
  Gemini \cite{team2023gemini} & Google & 500B & text & 32K & \Checkmark & \XSolidBrush & \XSolidBrush\\ 
  \hline
  CodeLlaMA \cite{codellama} & Meta AI & 13B & text \& code & 100K & \Checkmark & \Checkmark & \XSolidBrush\\ 
 \bottomrule
\end{tabular*}
\caption{Our study on LLMs understanding of decompiled binary vulnerable code: Vul-1 is a buffer overflow weakness in our defined \texttt{matrixmul} function, Vul-2 is a standard CWE-78 (OS Command Injection) in Juliet dataset. We compiled both of them to binary ELF files and analyze the decompiled code through LLMs. No LLM can find OS Command Injection intuitively from binary level but some of them can understand simple code (e.g. \texttt{matrixmul}) with synthetic weakness. NOTE: The 100k context window for CodeLlaMA requires intense computational power to process it.}
\label{table_genai}
\end{table*}

\begin{figure*}[!t]
\centering
\includegraphics[width=\linewidth]{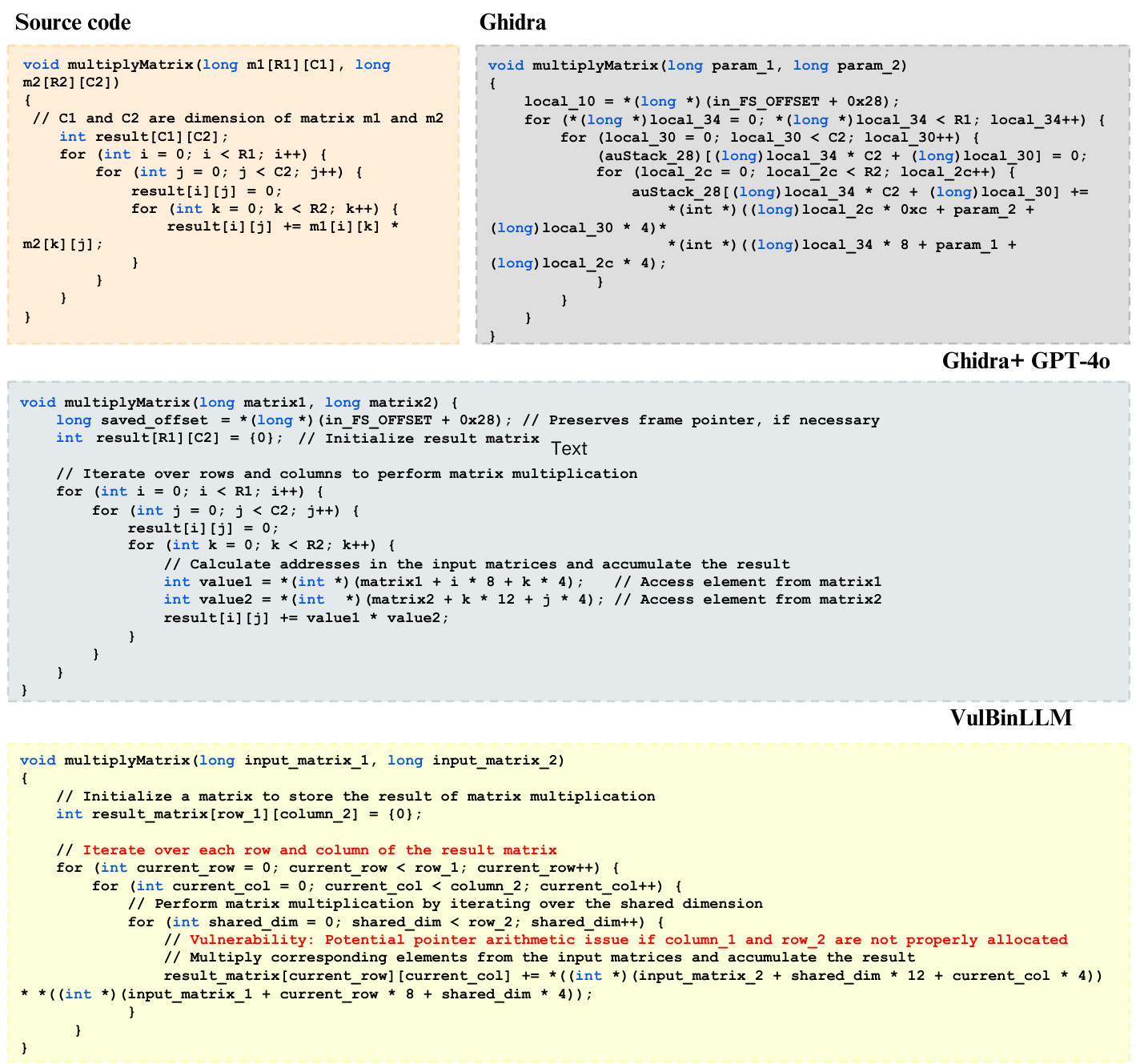}
\caption{An example of \texttt{Matrix Multiplication} decompilation output across different stages: the original source code, Ghidra, Ghidra enhanced with GPT-4o, and \sysname. The Ghidra decompilation provides a low-level representation with generic variable names and lacks context, making the functionality and security aspects harder to interpret. Ghidra + GPT-4o improves readability with meaningful variable names and clarifying comments. \sysname further augments the output by adding vulnerability-specific annotations, such as warnings about potential pointer arithmetic issues that could lead to buffer overflows or memory access vulnerabilities. This layered enhancement helps bridge the gap between decompilation and security analysis, making \sysname particularly beneficial for identifying and understanding vulnerabilities in binary code.}
\label{fig_decompile}
\end{figure*}

\footnotetext{For the buffer overflow weakness, we included the vulnerability by not setting the bound for the array}

\subsection{Design of \sysname} 
\label{method1}

\begin{figure*}[!t]
\centering
\includegraphics[width=\linewidth]{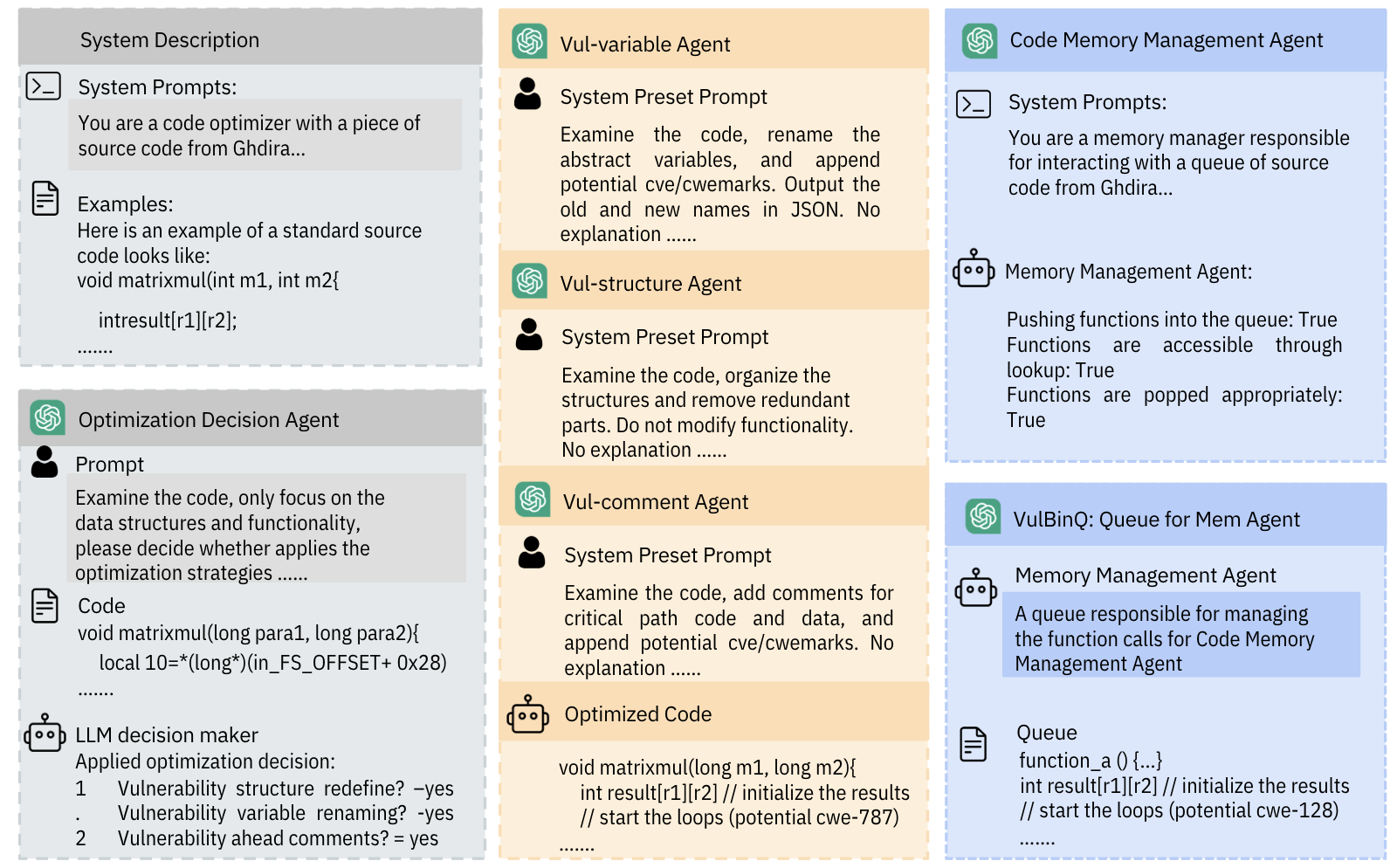}
\caption{\sysname-Decompiler overview. \sysname-decompiler includes an Optimization Decision Agent and three Action Agents (Vul-variable, Vul-struct, Vul-comment). After getting raw decompilation output from reverse engineering tool, \sysname-decompiler will perform an initial check on the grammar, functionality, and structures and decide what optimization decision will be made. Be sending requests to Action Agents, \sysname-decompiler will focus on renaming the variables and functions' names, reorganize the defined code structure, and critically, appending explanations on potentially vulnerability and functionalities attached the code. \sysname-decompiler can also support to learn examples code by in-context learning}
\label{fig_decompwork}
\end{figure*}

The overview of \sysname is illustrated in Figure \ref{fig_decompwork}. The key insight behind \sysname is to emulate the traditional binary analysis workflow while optimizing each step with LLM-powered enhancements. In conventional binary analysis, binaries are first disassembled and decompiled into source code or intermediate representations (IR) to then be analyzed using various static analysis techniques to detect vulnerabilities in the binary as the source code and IR represent the same functionality but are not easily understandable by a human. We integrate an LLM-powered workflow to assist in tradition binary analysis vulnerability detection workflow which allows for a better analysis of vulnerable binaries \sysname approaches these challenges using neural decompilation to recover high-level, vulnerability-related syntactic information from binary code, enabling LLMs to better understand program logic and identify potential vulnerabilities. We incorporate LLM models that analyze the decompiled code in a context-sensitive manner, considering the relationships between different code segments and the overall program structure, by utilizing an extended context window which serves as an archival storage for the analysis of vulnerabilities in functions within the source code. The archival storage is a SQL Database which allows the LLM Agent to store and fetch the summarized analysis of different functions within the source code file. We also emulate a queue within \sysname to ensure the coverage of all functions within a binary. 

\subsubsection{\sysname - Vulnerability Prominence}

Neural decompilation acts as a bridge between the low-level world of binary code and the high-level understanding of LLMs. By recovering source code from stripped binaries, it unlocks the potential for LLMs to analyze and comprehend the program's logic, paving the way for effective vulnerability detection. This recovered source code, however, is not merely a verbatim translation of the binary. It is optimized specifically for vulnerability detection, with embedding the key features and potential security flaws highlighted for LLMs to focus on.\ref{fig_decompile} This optimization process involves identifying and emphasizing code patterns that are commonly associated with vulnerabilities. These patterns can include dangerous function calls, such as those known to be susceptible to buffer overflows or injection attacks, as well as code constructs that often lead to memory corruption or logic errors. By highlighting these vulnerability-related features, the recovered source code becomes a more informative and targeted input for LLMs, guiding their analysis towards potential security flaws. Furthermore, the optimization process can involve incorporating contextual information into the recovered source code. This can include information about overall program structure, the relationships between different code segments, and the intended functionality of the program. By providing LLMs with this broader context, they can better understand the implications of specific code patterns and make more accurate vulnerability assessments. 
LLMs, with their ability to learn complex patterns and representations from data, can then analyze the code and identify potential vulnerabilities based on the highlighted features and the overall code structure. This approach combines the strengths of both neural decompilation and LLMs, enabling the detection of vulnerabilities in stripped binary files that were previously difficult to analyze with traditional methods. In essence, vulnerability feature optimized source code recovery transforms binary code into a format that is both understandable and actionable for LLMs. It bridges the gap between the low-level representation of binary code and the high-level reasoning capabilities of LLMs, enabling the efficient and effective detection of vulnerabilities in stripped binary files. Custom data types and user-defined classes are generally outside the scope of these tools, leading to inaccurate or incomplete representations. For instance, the decompiled output of a simple \textit{matrixMultiply()} function may display significant discrepancies in variable names and introduce overly complex or incorrect data structures. Furthermore, comments from the original source code are lost during decompilation since traditional reverse engineering tools cannot reconstruct them. Given our goal of using LLMs to analyze decompiled source code, preserving meaningful variable names, clear code structures, and informative comments is essential for enhancing the model’s comprehension and processing capability. Several recent approaches, such as DeGPT \cite{hu2024degpt} and ReSym \cite{xie2024resym}, leverage LLMs to improve decompilation outputs. However, these methods are not tailored for vulnerability detection, limiting their effectiveness in security-focused analysis. In order to make the decompilation efficient to be analyzed by LLMs, we use an LLM to first do a syntactical recovery of comment, structure and variable information from the decompiled code from RetDec which is then utilized by the detection agent with an extended context window to analyze the program state with respect to the syntactically descriptive code. In this work, we specifically focus on optimizing decompilation for identifying vulnerabilities alongside extending the LLMs capability to evaluate binaries, addressing the unique challenges in this domain.

\textbf{Prompt Engineering}. We utilized in-context learning and few-shot Chain-of-Thought (CoT) \cite{wei2022chain} and in-context-learning \cite{dong2022survey} prompting to enhance LLMs’ capability in identifying potential vulnerabilities. In-context learning enables LLMs to understand complex reasoning tasks by exposing them to examples of similar vulnerability patterns. This approach is particularly beneficial for vulnerability detection, as it allows the model to adapt to specific security weaknesses through example-based guidance. Few-shot CoT further decomposes complex reasoning into smaller, logical steps, prompting the LLMs to analyze code snippets by examining functionality, root causes, and potential impacts. This structured approach mirrors the systematic methods employed by security experts, allowing the model to connect multiple dimensions of a vulnerability. CoT is especially effective in uncovering hidden risks by providing a multi-faceted analysis aligned with expert practices.

\textbf{Construct Knowledge Documents}. Several datasets, such as Devign \cite{zhou2019devign}, BigVul \cite{he2023large}, and CodeSearchNet \cite{fan2020ac}, are commonly used to benchmark vulnerability detection performance. PairVul \cite{du2024vul} is a unique resource containing pairs of vulnerable and patched code samples. The memory management agent is connected to the VulBinQ, which is responsible for managing the functions that are to be analyzed by the LLM and various knowledge documents are created from the binary code to then be placed into the archived analysis data store allowing for a clearer representation of the binary code for the LLM. 

\section{Evaluation}
\label{eval}

In this section, we evaluate the performance of \sysname with an existing state-of-the-art approach to analyze each part's effectiveness.

\subsection{Implementation}
The implementation of \sysname is built upon the reverse engineering frameworks Ghidra \cite{ghidra} and \cite{retdec}. The \sysname decompilation utilizes the large language model GPT-4o to enhance vulnerability detection by carefully embedding the vulnerability information into the decompiled code. Below, we outline the implementation process. First, the binary is loaded and analyzed using Ghidra plugins. Ghidra facilitates the extraction of decompiled code, control flow graphs, import tables, stack frames, and other relevant information for subsequent analysis. Through the API of GPT-4.0 \cite{openai2024chatgpt}, \sysname identifies weaknesses in the binary and augments the output by (1) appending comments about potential vulnerabilities and functionality, (2) simplifying code structures, and (3) renaming variables for clarity and to make the vulnerability features more prominent. 

Following these enhancements, \sysname performs binary vulnerability classification using GPT-4o \cite{gpt4o}, leveraging vulnerability descriptions from MITRE for CWEs and then utilizes a extended context window approach with a shared contextual memory that allows for agents to reason about vulnerabilities in the decompiled code. Utilizing this agentic unbounded context window along with prompt templates and advanced prompt engineering techniques alongside a function queue, \sysname generates structured prompt sequences to guide GPT-4o in vulnerability inspection. GPT-4o processes code snippets in the context of these instructions, performing detailed analysis to identify potential vulnerabilities in the binary.

\subsection{Research questions and evaluation setup}


In order to address the research questions described above in section \ref{method}, we evaluate \sysname on binaries from the Juliet dataset, then compare the accuracy of the detected vulnerability on unstripped versions of these binaries. We also analyze how accuracy is affected with \sysname's extended context window approach for smaller binaries.

We evaluate the accuracy of \sysname's accuracy by the following methods:

\begin{itemize}
    \item How accurate is LLMs in detecting vulnerabilities in the stripped and unstipped binaries?
    \item Can we extend the LLM's capability to analyze binaries that far exceed it's context window?
    \item Can we utilize LLMs to detect vulnerabilities in the stripped synthetic test suits and compare it with state-of-the-art LLM powered tools?
\end{itemize}

In order to formulate our results we take two approaches for binary vulnerability analysis where the LLMs are prone to hallucinations and might require additional analysis for verification. We benchmark our system on the decompiled code from Juliet Test Suite.

\subsection{Results}
We evaluated \sysname on CWE classification on binary decompiled code. For the evaluation of \sysname, we utilized stripped Juliet Test Suite binaries. We use the following \textbf{Dataset:}

\begin{itemize}
    \item \textbf{Juliet Test Suite (v1.3) \cite{juliet}:} This C/C++ vulnerability test suite is organized by CWE categories and includes vulnerabilities relevant to our study (e.g., CWE-78, CWE-134, CWE-190, CWE-606). For evaluation, we compiled the test cases into binaries, removing debug information and symbol tables to simulate real-world scenarios. Test cases involving constant values were excluded, as LATTE targets vulnerabilities introduced by external inputs. Using GCC as our compiler, we generated over 20,000 binary samples. Table 2 summarizes the number of test cases used in our analysis.
    
\end{itemize}

\noindent\textbf{Evaluations on Synthetic Dataset:} For the evaluations on sythetic dataset (Juliet), we  compare \sysname against LATTE \cite{liu2023harnessing}, the state-of-the-art approach in utilizing LLMs for binary taint analysis. LATTE enhances vulnerability detection precision by integrating flow analysis with prompt-engineered LLM responses, ensuring consistency and reliability across multiple vulnerability categories but is limited to a certain class of CWEs that exist in binary files which are subject to be detected by binary taint analysis. We provide our results in Table \ref{table_latte} where we compare \sysname's abilities with LATTE.

\begin{table*}[ht]
\centering
\begin{tabular*}{\textwidth}{@{\extracolsep{\fill}} l l l l l l l l l}
\toprule
& \multicolumn{2}{c}{CWE-78 (960/960)$^+$} & \multicolumn{2}{c}{CWE-134 (1200/1200)$^+$} & \multicolumn{2}{c}{CWE-190 (2860/2860)$^+$} & \multicolumn{2}{c}{CWE-606 (240/240)$^+$} \\
\cmidrule(lr){2-3} \cmidrule(lr){4-5} \cmidrule(lr){6-7} \cmidrule(lr){8-9}
& LATTE & \sysname & LATTE & \sysname & LATTE & \sysname & LATTE & \sysname \\
\midrule
TP & 892  & 1055 & 1151 & 1345 & 1773 & 1725  & 210 & 1218\\
FN & 68   & 0 & 49   &  0 & 1087 &  0 & 30 & 0\\
TN & 960  & 4288 & 1102 & 3998 & 1779 &  3618 & 142 & 4125\\
FP & 0    & 191 & 98   & 14 & 1081 &  35 & 98 & 416\\
Accuracy & 96.46\% & \textbf{96.55\%}  & 93.88\% &  \textbf{99.74\% } & 62.1\%  & \textbf{99.34\%} & 73.33\% & \textbf{74.54\%}\\
Precision & \textbf{100\%} &  84.67\%  & 95.92\% &  \textbf{98.97\%}  & 52.04\% & \textbf{98.01\%} & 68.18\% & \textbf{92.78\%}\\
F1 Score & \textbf{96.33\%} &  91.70\%  & 93.99\% &  \textbf{99.48\%}  & 62.05\% & \textbf{98.99\%} & 74.24\% & \textbf{85.4\%}\\

\bottomrule
\end{tabular*}
\caption{Evaluation results of Vulnerability inspection based on Juliet Testsuite.}
\begin{tablenotes}
    \item[$+$] +:LATTE evaluation is done by embedding additional information alongside the code of the stripped binary to detect the CWE in the test suite.
    \item[$+$] +:The numbers in parentheses indicate the number of test cases (bad/good) for this vulnerability type.
    \item[$+$] +:The values for \sysname represents the number of files which contained various (bad/good) vulnerability types.
    \item[$+$] +:TN are all the files such that the vulnerability was in the prompt but was correctly  undetected by the LLM.
\end{tablenotes}
\label{table_latte}
\end{table*}

\begin{figure}[!t]
\centering
\includegraphics[width=\linewidth]{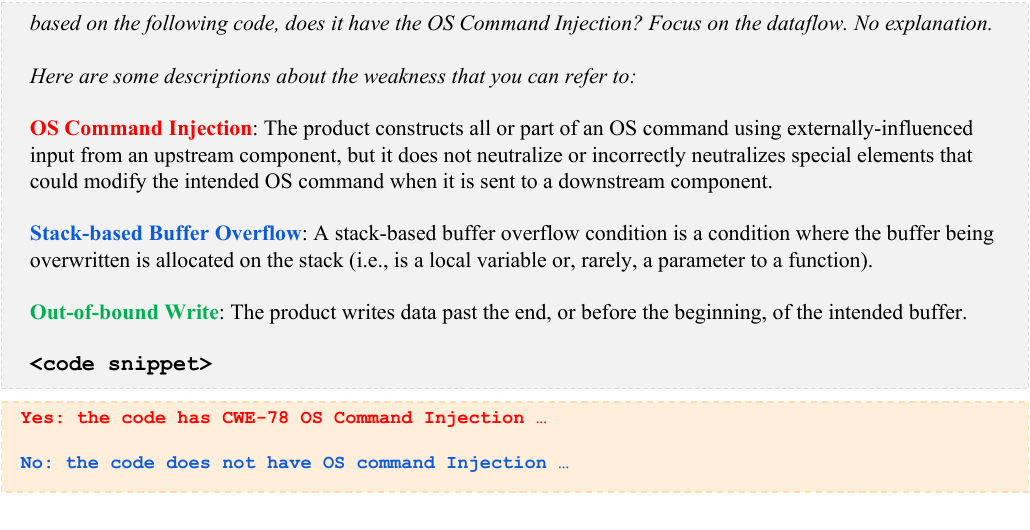}
\caption{An example of binary classification with CWE-78. The LLM is required to respond with \texttt{yes} or \texttt{no}, when asked if it is concentrating on the code flow rather than semantics. To avoid memorization  of LLMs in our special case, we append multiple CWEs (CWE-121: Stack Buffer Overflow, CWE-787: Out-of-bound Write)}
\label{fig_icl}
\end{figure}

\textbf{CWE Classification in \sysname}In \sysname's CWE classification task, we prompt GPT-4o to respond with the appropriate vulnerability (e.g., \texttt{CWE-78: OS Command Injection}) to determine the presence of a specific vulnerability in the optimized, neurally decompiled binary code. As shown in the Figure \ref{fig_icl}, we provide the model with a code snippet alongside a list of descriptions for potential weaknesses, such as OS Command Injection, Stack-based Buffer Overflow, and Out-of-bound Write. These descriptions are sourced directly from authoritative references like the MITRE CWE database, ensuring accuracy in the model's understanding of each vulnerability type. 

To minimize the risk of memorization, where the model might rely on learned patterns rather than performing genuine analysis, we include multiple CWEs in each query. By appending multiple vulnerability types and directing GPT-4o to focus on particular aspects—like data flow in this example—we guide the model to discern the nuanced characteristics of each potential weakness. This approach leverages extended context solutions with a stack that contains the decompiled source code functions to then combine contextual information with targeted prompt engineering, enabling \sysname to make informed vulnerability assessments within binary code.

\section{Discussions}

Despite the promising results demonstrated by \sysname, certain limitations remain First, due to the scarcity of binary vulnerability datasets, we evaluated \sysname on compiled data from the Juliet test suite. Although popular source code datasets such as BigVul, Devign, and REVEAL are often used in vulnerability research, they are challenging to apply in binary analysis due to inconsistent compilation environments and build dependencies. For example, Devign consists of complex projects requiring specific Makefiles for compilation, while BigVul includes diverse projects with inconsistent toolchains for building vulnerable code. There's an ongoing effort to improve the neural decompilation process for binaries and to optimize the decompiled code to recover syntactic details out of the semantic information that is created by the decompiler \cite{tan2024llm4decompile} \cite{xie2024resym}. Another approach that can be pursued to extend the LLM's capability to analyze a binary through LLMs without having to go through the decompilation process is to directly detect vulnerabilities on the assembly level. The critical issue with such a process it the high syntactic similarity between two binaries for a single architecture. In our evaluations we evaluated various similarity metrics including cosine similarity and Levenshtein distance between two different CWE examples in X86 assembly representation from the Juliet Test Suite, we detected the cosine similarity to be approximately 98\% for most of the CWEs in there even though the semantic information is very different for each of those files. It occurs due to the similarity among the two binaries in their assembly format referring to the same registers and the same architecture-specific assembly instructions. Another important field of research is to extend the LLM's capability to reason about complex tasks, especially vulnerability reports. The justification in CVE reports is provided by either a clear definition of the vulnerability or an exploit method that constitutes as a piece of sufficient evidence to show the existence of a vulnerability in the source code. However, in a binary describing a vulnerability would require the availability of an attack method that is then run to ensure the availability of such vulnerability in a binary. We consider the explainability of the reasoning of the vulnerability to be out of scope for this paper and a future direction. We also lack a comprehensive classification model for the definition of vulnerabilities in the binary level. For source code vulnerability classification, we have CVE and CWE definitions which provide both a fine and coarse-grained classification for vulnerabilities which is not defined for architecture-specific binary vulnerabilities. Another future direction for \sysname is to explore the areas of formal specification, probabilistic inference mechanisms, retrieval augmented generation, small language model alongside a descriptive analysis of architecture specific vulnerability definition as future directions to advance the binary analysis ecosystem and to introduce LLM-like solutions to assign security experts to potentially analyze binaries for vulnerabilities and malware.

\section{Related Work}
\label{related}

\subsection{Semantic Recovery from Binary files}

While LLMs have shown promise in various software engineering tasks, their application in binary analysis has primarily focused on syntactic regeneration of code. Previous research works \cite{xie2024resym} \cite{hu2024degpt} has explored using LLMs to generate human-readable representations of binary code, such as recovering function names or providing code summaries. These approaches leverage the LLMs' ability to learn patterns and structures in code to reconstruct higher-level representations from the low-level binary code. However, these efforts primarily aim to improve code understandability for human analysts rather than directly detecting vulnerabilities. They focus on regenerating the syntactic structure of the code without necessarily delving into the identification of potential security flaws. This limitation stems from the inherent challenges of binary code analysis, where the absence of high-level abstractions and semantic information makes it difficult to extract vulnerability-related features directly from the binary.

\subsection{Reverse Engineering and Program Analysis for Vulnerability Detection}

Reverse engineering is the process of analyzing and understanding the design, structure, and functionality of a code snippet by working backward from its final form. It is a critical technique in software engineering particularly within binary analysis. Traditional binary analysis is involved in decompilation, which is a significant portion in reverse engineering. Reverse engineering has been adopted in many program analysis domains, e.g. vulnerability assessment \cite{angelakopoulos2023firmsolo, cochard2022investigating, he2024code, kim2022revisiting, pei2022learning, vadayath2022arbiter}, malware analysis \cite{chen2021selectivetaint, gao2018vulseeker, wang2023can, you2020pmp}, software repair \cite{ullah2024llms, li2023cctest, pearce2023examining, peng2024domain}, and code optimization \cite{maheswaranathan2021reverse}. 

Due to the inherent difficulty and opacity of binary code, prior research has focused on the readability and maintainability. Similarity analysis \cite{wang2022jtrans, xu2023improving}, memory analysis \cite{pei2022neudep}, assembly-to-code translation \cite{ahmad2023unsupervised, gouicem2022risotto}, function identification \cite{kim2023funprobe}. Decompilation optimization is an active research area since it combines reverse engineering and machine learning. David et al. \cite{david2020neural} leverage LSTM \cite{hochreiter1997long} to predict variable names in stripped binaries. TIE \cite{lee2011tie}, Retyped \cite{noonan2016polymorphic}, and OSPREY \cite{zhang2021osprey} focus on variable type recovery. Direct \cite{nitin2021direct} and DIRTY \cite{chen2022augmenting} have exploited Transformer-based methods for recovering variable names from decompiled code. More recently, DeGPT \cite{hu2024degpt} and Resym \cite{xie2024resym} leverage LLMs to identify, and recover code that is readable and similar to ground-truth original code. \sysname, on the other hand, emphasizes on annotating potential vulnerability information while simplifying code structures and renaming variable names, while prior work \cite{he2018debin, chen2020cati} employ ML-based approaches to predict debugging information from stripped binaries. \sysname can generate more concrete vulnerability comments to provide appropriate information for the LLM to analyze it for vulnerability detection.


\subsection{Learning-based Vulnerability Detection}
\label{learning_vul_method}

Learning-based approaches have made substantial progress in vulnerability detection \cite{zhou2019devign, mirsky2023vulchecker, chakraborty2021deep, steenhoek2024dataflow, hanif2022vulberta, fu2022linevul}, employing various methods such as graph representations and large language models to analyze code.
Graph Neural Network (GNN)-based techniques \cite{zhou2019devign, mirsky2023vulchecker, tang2023csgvd} represent code snippets as graph-based structures, such as Abstract Syntax Trees (ASTs) and Control Flow Graphs (CFGs). By utilizing GNN on these graph based structured representations, It allows for a better analysis for vulnerabilities in the representation of control flow graphs, call graphs and code property graphs. However, GNN-based methods face challenges in effectively distinguishing vulnerabilities within lexically-similar but semantically-different code pairs, as they are limited by their reliance on structural information alone, often lacking deeper semantic understanding required to identify subtle vulnerabilities. \cite{du2024vul}


Other learning-based approaches have explored the use of \\transformer-based models, where Large Language Models (LLMs) have been employed to detect vulnerabilities by leveraging large-scale pre-training on code \cite{liu2024pre, peng2023ptlvd}. Although LLMs have shown promise in generating syntactically correct and contextually relevant code, their effectiveness in identifying vulnerabilities is limited when they encounter complex low-level vulnerabilities or intricate program flows, which require precise interpretation of code semantics and memory handling \cite{zhou2024large}.

In addition, traditional static analysis tools, such as Cppcheck, offer rule-based analysis to detect code issues, including potential vulnerabilities. While these tools are efficient and interpretable, they often suffer from high false-positive rates and lack the flexibility to detect novel or context-dependent vulnerabilities, as they cannot generalize beyond predefined rules.

PLM-based Vulnerability Detection involves fine-tuning pre-trained language models (PLMs) on vulnerability detection datasets. In this approach, code snippets are tokenized and processed by PLMs (e.g., RoBERTa \cite{liu2019roberta}), which act as encoders. The PLMs extract semantic features from the code, which are then used in binary classification to determine the presence of vulnerabilities. This method leverages the language understanding capabilities of PLMs to analyze code textually, making it suitable for detecting vulnerabilities that are evident from code semantics.

LLM-based Vulnerability Detection utilize LLMs through either prompt engineering or fine-tuning. Prompt engineering strategies, such as Chain-of-Thought (CoT) \cite{wei2022chain} reasoning and few-shot learning, enable LLMs to detect vulnerabilities more accurately without modifying the original model parameters. In contrast, the fine-tuning approach involves training LLMs on vulnerability detection datasets, allowing the models to learn specific features of vulnerable code by updating their parameters. This approach takes advantage of LLMs’ strong contextual understanding, making it well-suited for identifying complex vulnerabilities across diverse codebases.

\subsection{Code Large Language Models (CodeLLMs)}

With the power of Transformer architecture \cite{vaswani2017attention}, modern large language models (LLMs) show great potential in code related tasks. Specifically, for encoder-only models (e.g., BERT \cite{devlin2018bert}, RoBERTa \cite{liu2019roberta}), Code-BERT \cite{feng2020codebert} learns from massive source code and natural language descriptions to show promising results in code search, completion, and summarizations. GraphBert \cite{zhang2020graph} can learn representations from graph-structured code to leverage structural information. There are also followed-up work based on them Code-BERT \cite{feng2020codebert}. Furthermore, for decoder-only models (e.g., GPTs \cite{brown2020language, radford2018improving}), GPT-4 with canvas \cite{gpt4}, CodeLLaMA series \cite{roziere2023code, touvron2023llama, dubey2024llama}, Claude Sonnet series \cite{anthropic2024claude}, Mistral \cite{mistral}\& Codestral \cite{codestral}, DeepSeek Code \cite{guo2024deepseek}, and Gemini \cite{team2023gemini} enhance the capabilities of understanding code related tasks specifically. Recently, more emerging models like Codestral Mamba \cite{codestralmamba} offer the advantage of linear time inference \cite{gu2023mamba, dao2024transformers} and the theoretical ability to model sequences of infinite length. However, those Code LLMs are focus on code generation, debugging, and summarization \cite{ahmad2021unified, fried2022incoder, guo2024deepseek, li2023starcoder, nijkamp2022codegen, roziere2023code, wang2021codet5}. Our work focuses on code analysis especially in vulnerability or weakness.

There are prior works applying LLMs in code understanding \cite{nam2024using, chen2023teaching, madaan2024self, olausson2023demystifying, scheurer2023training}, software fuzzing \cite{yu2024llm, kang2023large, deng2023large}, natural language alignment \cite{neelakantan2022text, geng2023empirical, muennighoff2023octopack, xie2023impact}, vulnerability repair \cite{xia2023automated, wu2023effective, steenhoek2023empirical, jiang2023impact, du2024vul}. There are few works in applying LLMs in binary analysis. LATTE \cite{liu2023harnessing} is the most relevant works. However, it focuses on specific taint analysis. Thus, we treat is as a complement to our work. Recent work in binary-related work \cite{su2024codeart, yang2023asteria, li2023vulanalyzer, su2024source, luo2023vulhawk} are all based on traditional machine leaning methods rather than LLMs. For example, CodeArt \cite{su2024codeart} pretrains a BERT-like model on binary functions through explicit attention mechanisms. VulHawk leverages an intermediate-representation by RoBERTa with code reuse similarity. VulANalyzeR exploits Gragh Convolution and attention mechanisms to classify vulnerabilities from Control Flow Graphs. Note that DeBinVul is an orthogonal work to ours. They fine-tuned a Code LLM to detect vulnerabilities from decompiled code with self-built datasets. On the other hand, \sysname is able to detect vulnerabilities without fine-tuning. Further, \sysname is easy to scale to multiple programming languages and have a better generalization due to its flexibility.

\section{Conclusions}
This paper introduces Vul-BinLLM, an LLM-based framework for binary vulnerability detection that integrates decompilation optimization utilizing neural decompilation and extended context window and memory management capabilities to enable  vulnerability analysis with binaries. The results demonstrate the potential of LLMs to address longstanding challenges in binary vulnerability detection, paving the way for more scalable and secure software systems.

\bibliographystyle{ACM-Reference-Format}
\bibliography{sample-base}


\begin{thebibliography}{118}


\ifx \showCODEN    \undefined \def \showCODEN     #1{\unskip}     \fi
\ifx \showDOI      \undefined \def \showDOI       #1{#1}\fi
\ifx \showISBNx    \undefined \def \showISBNx     #1{\unskip}     \fi
\ifx \showISBNxiii \undefined \def \showISBNxiii  #1{\unskip}     \fi
\ifx \showISSN     \undefined \def \showISSN      #1{\unskip}     \fi
\ifx \showLCCN     \undefined \def \showLCCN      #1{\unskip}     \fi
\ifx \shownote     \undefined \def \shownote      #1{#1}          \fi
\ifx \showarticletitle \undefined \def \showarticletitle #1{#1}   \fi
\ifx \showURL      \undefined \def \showURL       {\relax}        \fi
\providecommand\bibfield[2]{#2}
\providecommand\bibinfo[2]{#2}
\providecommand\natexlab[1]{#1}
\providecommand\showeprint[2][]{arXiv:#2}

\bibitem[cod(2023)]%
        {codellama}
 \bibinfo{year}{2023}\natexlab{}.
\newblock \bibinfo{booktitle}{\emph{CodeLlaMA by Meta AI}}.
\newblock
\urldef\tempurl%
\url{https://ai.meta.com/blog/code-llama-large-language-model-coding/}
\showURL{%
\tempurl}


\bibitem[gpt(2023)]%
        {gpt4}
 \bibinfo{year}{2023}\natexlab{}.
\newblock \bibinfo{booktitle}{\emph{GPT-4 by OpenAI}}.
\newblock
\urldef\tempurl%
\url{https://openai.com/research/gpt-4}
\showURL{%
\tempurl}


\bibitem[big(2024)]%
        {big_vul}
 \bibinfo{year}{2024}\natexlab{}.
\newblock \bibinfo{booktitle}{\emph{Big-Vul Dataset}}.
\newblock
\urldef\tempurl%
\url{https://huggingface.co/datasets/bstee615/bigvul}
\showURL{%
\tempurl}


\bibitem[dev(2024)]%
        {devign}
 \bibinfo{year}{2024}\natexlab{}.
\newblock \bibinfo{booktitle}{\emph{Devign Dataset}}.
\newblock
\urldef\tempurl%
\url{https://github.com/epicosy/devign}
\showURL{%
\tempurl}


\bibitem[ghi(2024)]%
        {ghidra}
 \bibinfo{year}{2024}\natexlab{}.
\newblock \bibinfo{booktitle}{\emph{Ghidra}}.
\newblock
\urldef\tempurl%
\url{https://ghidra-sre.org/}
\showURL{%
\tempurl}


\bibitem[ida(2024)]%
        {idapro}
 \bibinfo{year}{2024}\natexlab{}.
\newblock \bibinfo{booktitle}{\emph{IDA Pro}}.
\newblock
\urldef\tempurl%
\url{https://hex-rays.com/ida-pro}
\showURL{%
\tempurl}


\bibitem[jul(2024)]%
        {juliet_c_cpp}
 \bibinfo{year}{2024}\natexlab{}.
\newblock \bibinfo{booktitle}{\emph{Juliet Test Suite for C/C++ and Java}}.
\newblock
\urldef\tempurl%
\url{https://samate.nist.gov/SARD/testsuite.php}
\showURL{%
\tempurl}


\bibitem[mvd(2024)]%
        {mvd}
 \bibinfo{year}{2024}\natexlab{}.
\newblock \bibinfo{booktitle}{\emph{Microsoft Vulnerability Dataset (MVD)}}.
\newblock
\urldef\tempurl%
\url{https://github.com/microsoft/MS-MVD}
\showURL{%
\tempurl}


\bibitem[ret(2024)]%
        {retdec}
 \bibinfo{year}{2024}\natexlab{}.
\newblock \bibinfo{booktitle}{\emph{RetDec}}.
\newblock
\urldef\tempurl%
\url{https://github.com/avast/retdec}
\showURL{%
\tempurl}


\bibitem[rev(2024)]%
        {reveal}
 \bibinfo{year}{2024}\natexlab{}.
\newblock \bibinfo{booktitle}{\emph{REVEAL Dataset}}.
\newblock
\urldef\tempurl%
\url{https://github.com/VulDetProject/ReVeal}
\showURL{%
\tempurl}


\bibitem[sar(2024)]%
        {sard}
 \bibinfo{year}{2024}\natexlab{}.
\newblock \bibinfo{booktitle}{\emph{Software Assurance Reference Dataset (SARD)}}.
\newblock
\urldef\tempurl%
\url{https://samate.nist.gov/SARD/}
\showURL{%
\tempurl}


\bibitem[vul(2024)]%
        {vul4j}
 \bibinfo{year}{2024}\natexlab{}.
\newblock \bibinfo{booktitle}{\emph{Vul4J Dataset}}.
\newblock
\urldef\tempurl%
\url{https://github.com/tuhh-softsec/vul4j}
\showURL{%
\tempurl}


\bibitem[Ahmad and Luo(2023)]%
        {ahmad2023unsupervised}
\bibfield{author}{\bibinfo{person}{Iftakhar Ahmad} {and} \bibinfo{person}{Lannan Luo}.} \bibinfo{year}{2023}\natexlab{}.
\newblock \showarticletitle{Unsupervised Binary Code Translation with Application to Code Clone Detection and Vulnerability Discovery}. In \bibinfo{booktitle}{\emph{Findings of the Association for Computational Linguistics: EMNLP 2023}}. \bibinfo{pages}{14581--14592}.
\newblock


\bibitem[Ahmad et~al\mbox{.}(2021)]%
        {ahmad2021unified}
\bibfield{author}{\bibinfo{person}{Wasi~Uddin Ahmad}, \bibinfo{person}{Saikat Chakraborty}, \bibinfo{person}{Baishakhi Ray}, {and} \bibinfo{person}{Kai-Wei Chang}.} \bibinfo{year}{2021}\natexlab{}.
\newblock \showarticletitle{Unified pre-training for program understanding and generation}.
\newblock \bibinfo{journal}{\emph{arXiv preprint arXiv:2103.06333}} (\bibinfo{year}{2021}).
\newblock


\bibitem[AI(2024a)]%
        {codestral}
\bibfield{author}{\bibinfo{person}{Mistral AI}.} \bibinfo{year}{2024}\natexlab{a}.
\newblock \bibinfo{booktitle}{\emph{Codestral}}.
\newblock
\urldef\tempurl%
\url{https://mistral.ai/news/codestral/}
\showURL{%
\tempurl}


\bibitem[AI(2024b)]%
        {codestralmamba}
\bibfield{author}{\bibinfo{person}{Mistral AI}.} \bibinfo{year}{2024}\natexlab{b}.
\newblock \bibinfo{booktitle}{\emph{Codestral Mamba}}.
\newblock
\urldef\tempurl%
\url{https://mistral.ai/news/codestral-mamba/}
\showURL{%
\tempurl}


\bibitem[AI(2024c)]%
        {mistral}
\bibfield{author}{\bibinfo{person}{Mistral AI}.} \bibinfo{year}{2024}\natexlab{c}.
\newblock \bibinfo{booktitle}{\emph{Mixture of Experts Models}}.
\newblock
\urldef\tempurl%
\url{https://mistral.ai/news/mixtral-of-experts/}
\showURL{%
\tempurl}


\bibitem[Angelakopoulos et~al\mbox{.}(2023)]%
        {angelakopoulos2023firmsolo}
\bibfield{author}{\bibinfo{person}{Ioannis Angelakopoulos}, \bibinfo{person}{Gianluca Stringhini}, {and} \bibinfo{person}{Manuel Egele}.} \bibinfo{year}{2023}\natexlab{}.
\newblock \showarticletitle{$\{$FirmSolo$\}$: Enabling dynamic analysis of binary Linux-based $\{$IoT$\}$ kernel modules}. In \bibinfo{booktitle}{\emph{32nd USENIX Security Symposium (USENIX Security 23)}}. \bibinfo{pages}{5021--5038}.
\newblock


\bibitem[anthropic(2024)]%
        {anthropic2024claude}
\bibfield{author}{\bibinfo{person}{anthropic}.} \bibinfo{year}{2024}\natexlab{}.
\newblock \bibinfo{booktitle}{\emph{Claude}}.
\newblock
\urldef\tempurl%
\url{https://www.anthropic.com/claude}
\showURL{%
\tempurl}


\bibitem[Brown(2020)]%
        {brown2020language}
\bibfield{author}{\bibinfo{person}{Tom~B Brown}.} \bibinfo{year}{2020}\natexlab{}.
\newblock \showarticletitle{Language models are few-shot learners}.
\newblock \bibinfo{journal}{\emph{arXiv preprint arXiv:2005.14165}} (\bibinfo{year}{2020}).
\newblock


\bibitem[Chakraborty et~al\mbox{.}(2021)]%
        {chakraborty2021deep}
\bibfield{author}{\bibinfo{person}{Saikat Chakraborty}, \bibinfo{person}{Rahul Krishna}, \bibinfo{person}{Yangruibo Ding}, {and} \bibinfo{person}{Baishakhi Ray}.} \bibinfo{year}{2021}\natexlab{}.
\newblock \showarticletitle{Deep learning based vulnerability detection: Are we there yet?}
\newblock \bibinfo{journal}{\emph{IEEE Transactions on Software Engineering}} \bibinfo{volume}{48}, \bibinfo{number}{9} (\bibinfo{year}{2021}), \bibinfo{pages}{3280--3296}.
\newblock


\bibitem[Chen et~al\mbox{.}(2020)]%
        {chen2020cati}
\bibfield{author}{\bibinfo{person}{Ligeng Chen}, \bibinfo{person}{Zhongling He}, {and} \bibinfo{person}{Bing Mao}.} \bibinfo{year}{2020}\natexlab{}.
\newblock \showarticletitle{Cati: Context-assisted type inference from stripped binaries}. In \bibinfo{booktitle}{\emph{2020 50th Annual IEEE/IFIP International Conference on Dependable Systems and Networks (DSN)}}. IEEE, \bibinfo{pages}{88--98}.
\newblock


\bibitem[Chen et~al\mbox{.}(2022)]%
        {chen2022augmenting}
\bibfield{author}{\bibinfo{person}{Qibin Chen}, \bibinfo{person}{Jeremy Lacomis}, \bibinfo{person}{Edward~J Schwartz}, \bibinfo{person}{Claire Le~Goues}, \bibinfo{person}{Graham Neubig}, {and} \bibinfo{person}{Bogdan Vasilescu}.} \bibinfo{year}{2022}\natexlab{}.
\newblock \showarticletitle{Augmenting decompiler output with learned variable names and types}. In \bibinfo{booktitle}{\emph{31st USENIX Security Symposium (USENIX Security 22)}}. \bibinfo{pages}{4327--4343}.
\newblock


\bibitem[Chen et~al\mbox{.}(2021)]%
        {chen2021selectivetaint}
\bibfield{author}{\bibinfo{person}{Sanchuan Chen}, \bibinfo{person}{Zhiqiang Lin}, {and} \bibinfo{person}{Yinqian Zhang}.} \bibinfo{year}{2021}\natexlab{}.
\newblock \showarticletitle{$\{$SelectiveTaint$\}$: Efficient Data Flow Tracking With Static Binary Rewriting}. In \bibinfo{booktitle}{\emph{30th USENIX Security Symposium (USENIX Security 21)}}. \bibinfo{pages}{1665--1682}.
\newblock


\bibitem[Chen et~al\mbox{.}(2023)]%
        {chen2023teaching}
\bibfield{author}{\bibinfo{person}{Xinyun Chen}, \bibinfo{person}{Maxwell Lin}, \bibinfo{person}{Nathanael Sch{\"a}rli}, {and} \bibinfo{person}{Denny Zhou}.} \bibinfo{year}{2023}\natexlab{}.
\newblock \showarticletitle{Teaching large language models to self-debug}.
\newblock \bibinfo{journal}{\emph{arXiv preprint arXiv:2304.05128}} (\bibinfo{year}{2023}).
\newblock


\bibitem[Cochard et~al\mbox{.}(2022)]%
        {cochard2022investigating}
\bibfield{author}{\bibinfo{person}{Victor Cochard}, \bibinfo{person}{Damian Pfammatter}, \bibinfo{person}{Chi~Thang Duong}, {and} \bibinfo{person}{Mathias Humbert}.} \bibinfo{year}{2022}\natexlab{}.
\newblock \showarticletitle{Investigating graph embedding methods for cross-platform binary code similarity detection}. In \bibinfo{booktitle}{\emph{2022 IEEE 7th European Symposium on Security and Privacy (EuroS\&P)}}. IEEE, \bibinfo{pages}{60--73}.
\newblock


\bibitem[Dao and Gu(2024)]%
        {dao2024transformers}
\bibfield{author}{\bibinfo{person}{Tri Dao} {and} \bibinfo{person}{Albert Gu}.} \bibinfo{year}{2024}\natexlab{}.
\newblock \showarticletitle{Transformers are SSMs: Generalized models and efficient algorithms through structured state space duality}.
\newblock \bibinfo{journal}{\emph{arXiv preprint arXiv:2405.21060}} (\bibinfo{year}{2024}).
\newblock


\bibitem[David et~al\mbox{.}(2020)]%
        {david2020neural}
\bibfield{author}{\bibinfo{person}{Yaniv David}, \bibinfo{person}{Uri Alon}, {and} \bibinfo{person}{Eran Yahav}.} \bibinfo{year}{2020}\natexlab{}.
\newblock \showarticletitle{Neural reverse engineering of stripped binaries using augmented control flow graphs}.
\newblock \bibinfo{journal}{\emph{Proceedings of the ACM on Programming Languages}} \bibinfo{volume}{4}, \bibinfo{number}{OOPSLA} (\bibinfo{year}{2020}), \bibinfo{pages}{1--28}.
\newblock


\bibitem[Deng et~al\mbox{.}(2023)]%
        {deng2023large}
\bibfield{author}{\bibinfo{person}{Yinlin Deng}, \bibinfo{person}{Chunqiu~Steven Xia}, \bibinfo{person}{Haoran Peng}, \bibinfo{person}{Chenyuan Yang}, {and} \bibinfo{person}{Lingming Zhang}.} \bibinfo{year}{2023}\natexlab{}.
\newblock \showarticletitle{Large language models are zero-shot fuzzers: Fuzzing deep-learning libraries via large language models}. In \bibinfo{booktitle}{\emph{Proceedings of the 32nd ACM SIGSOFT international symposium on software testing and analysis}}. \bibinfo{pages}{423--435}.
\newblock


\bibitem[Devlin(2018)]%
        {devlin2018bert}
\bibfield{author}{\bibinfo{person}{Jacob Devlin}.} \bibinfo{year}{2018}\natexlab{}.
\newblock \showarticletitle{Bert: Pre-training of deep bidirectional transformers for language understanding}.
\newblock \bibinfo{journal}{\emph{arXiv preprint arXiv:1810.04805}} (\bibinfo{year}{2018}).
\newblock


\bibitem[Dong et~al\mbox{.}(2022)]%
        {dong2022survey}
\bibfield{author}{\bibinfo{person}{Qingxiu Dong}, \bibinfo{person}{Lei Li}, \bibinfo{person}{Damai Dai}, \bibinfo{person}{Ce Zheng}, \bibinfo{person}{Jingyuan Ma}, \bibinfo{person}{Rui Li}, \bibinfo{person}{Heming Xia}, \bibinfo{person}{Jingjing Xu}, \bibinfo{person}{Zhiyong Wu}, \bibinfo{person}{Tianyu Liu}, {et~al\mbox{.}}} \bibinfo{year}{2022}\natexlab{}.
\newblock \showarticletitle{A survey on in-context learning}.
\newblock \bibinfo{journal}{\emph{arXiv preprint arXiv:2301.00234}} (\bibinfo{year}{2022}).
\newblock


\bibitem[Du et~al\mbox{.}(2024)]%
        {du2024vul}
\bibfield{author}{\bibinfo{person}{Xueying Du}, \bibinfo{person}{Geng Zheng}, \bibinfo{person}{Kaixin Wang}, \bibinfo{person}{Jiayi Feng}, \bibinfo{person}{Wentai Deng}, \bibinfo{person}{Mingwei Liu}, \bibinfo{person}{Bihuan Chen}, \bibinfo{person}{Xin Peng}, \bibinfo{person}{Tao Ma}, {and} \bibinfo{person}{Yiling Lou}.} \bibinfo{year}{2024}\natexlab{}.
\newblock \showarticletitle{Vul-RAG: Enhancing LLM-based Vulnerability Detection via Knowledge-level RAG}.
\newblock \bibinfo{journal}{\emph{arXiv preprint arXiv:2406.11147}} (\bibinfo{year}{2024}).
\newblock


\bibitem[Dubey et~al\mbox{.}(2024)]%
        {dubey2024llama}
\bibfield{author}{\bibinfo{person}{Abhimanyu Dubey}, \bibinfo{person}{Abhinav Jauhri}, \bibinfo{person}{Abhinav Pandey}, \bibinfo{person}{Abhishek Kadian}, \bibinfo{person}{Ahmad Al-Dahle}, \bibinfo{person}{Aiesha Letman}, \bibinfo{person}{Akhil Mathur}, \bibinfo{person}{Alan Schelten}, \bibinfo{person}{Amy Yang}, \bibinfo{person}{Angela Fan}, {et~al\mbox{.}}} \bibinfo{year}{2024}\natexlab{}.
\newblock \showarticletitle{The llama 3 herd of models}.
\newblock \bibinfo{journal}{\emph{arXiv preprint arXiv:2407.21783}} (\bibinfo{year}{2024}).
\newblock


\bibitem[Fan et~al\mbox{.}(2020)]%
        {fan2020ac}
\bibfield{author}{\bibinfo{person}{Jiahao Fan}, \bibinfo{person}{Yi Li}, \bibinfo{person}{Shaohua Wang}, {and} \bibinfo{person}{Tien~N Nguyen}.} \bibinfo{year}{2020}\natexlab{}.
\newblock \showarticletitle{AC/C++ code vulnerability dataset with code changes and CVE summaries}. In \bibinfo{booktitle}{\emph{Proceedings of the 17th International Conference on Mining Software Repositories}}. \bibinfo{pages}{508--512}.
\newblock


\bibitem[Feng et~al\mbox{.}(2020)]%
        {feng2020codebert}
\bibfield{author}{\bibinfo{person}{Zhangyin Feng}, \bibinfo{person}{Daya Guo}, \bibinfo{person}{Duyu Tang}, \bibinfo{person}{Nan Duan}, \bibinfo{person}{Xiaocheng Feng}, \bibinfo{person}{Ming Gong}, \bibinfo{person}{Linjun Shou}, \bibinfo{person}{Bing Qin}, \bibinfo{person}{Ting Liu}, \bibinfo{person}{Daxin Jiang}, {et~al\mbox{.}}} \bibinfo{year}{2020}\natexlab{}.
\newblock \showarticletitle{Codebert: A pre-trained model for programming and natural languages}.
\newblock \bibinfo{journal}{\emph{arXiv preprint arXiv:2002.08155}} (\bibinfo{year}{2020}).
\newblock


\bibitem[Fried et~al\mbox{.}(2022)]%
        {fried2022incoder}
\bibfield{author}{\bibinfo{person}{Daniel Fried}, \bibinfo{person}{Armen Aghajanyan}, \bibinfo{person}{Jessy Lin}, \bibinfo{person}{Sida Wang}, \bibinfo{person}{Eric Wallace}, \bibinfo{person}{Freda Shi}, \bibinfo{person}{Ruiqi Zhong}, \bibinfo{person}{Wen-tau Yih}, \bibinfo{person}{Luke Zettlemoyer}, {and} \bibinfo{person}{Mike Lewis}.} \bibinfo{year}{2022}\natexlab{}.
\newblock \showarticletitle{Incoder: A generative model for code infilling and synthesis}.
\newblock \bibinfo{journal}{\emph{arXiv preprint arXiv:2204.05999}} (\bibinfo{year}{2022}).
\newblock


\bibitem[Fu and Tantithamthavorn(2022)]%
        {fu2022linevul}
\bibfield{author}{\bibinfo{person}{Michael Fu} {and} \bibinfo{person}{Chakkrit Tantithamthavorn}.} \bibinfo{year}{2022}\natexlab{}.
\newblock \showarticletitle{Linevul: A transformer-based line-level vulnerability prediction}. In \bibinfo{booktitle}{\emph{Proceedings of the 19th International Conference on Mining Software Repositories}}. \bibinfo{pages}{608--620}.
\newblock


\bibitem[Gao et~al\mbox{.}(2018)]%
        {gao2018vulseeker}
\bibfield{author}{\bibinfo{person}{Jian Gao}, \bibinfo{person}{Xin Yang}, \bibinfo{person}{Ying Fu}, \bibinfo{person}{Yu Jiang}, {and} \bibinfo{person}{Jiaguang Sun}.} \bibinfo{year}{2018}\natexlab{}.
\newblock \showarticletitle{Vulseeker: A semantic learning based vulnerability seeker for cross-platform binary}. In \bibinfo{booktitle}{\emph{Proceedings of the 33rd ACM/IEEE International Conference on Automated Software Engineering}}. \bibinfo{pages}{896--899}.
\newblock


\bibitem[Geng et~al\mbox{.}(2023)]%
        {geng2023empirical}
\bibfield{author}{\bibinfo{person}{Mingyang Geng}, \bibinfo{person}{Shangwen Wang}, \bibinfo{person}{Dezun Dong}, \bibinfo{person}{Haotian Wang}, \bibinfo{person}{Ge Li}, \bibinfo{person}{Zhi Jin}, \bibinfo{person}{Xiaoguang Mao}, {and} \bibinfo{person}{Xiangke Liao}.} \bibinfo{year}{2023}\natexlab{}.
\newblock \showarticletitle{An empirical study on using large language models for multi-intent comment generation}.
\newblock \bibinfo{journal}{\emph{arXiv preprint arXiv:2304.11384}} (\bibinfo{year}{2023}).
\newblock


\bibitem[Gouicem et~al\mbox{.}(2022)]%
        {gouicem2022risotto}
\bibfield{author}{\bibinfo{person}{Redha Gouicem}, \bibinfo{person}{Dennis Sprokholt}, \bibinfo{person}{Jasper Ruehl}, \bibinfo{person}{Rodrigo~CO Rocha}, \bibinfo{person}{Tom Spink}, \bibinfo{person}{Soham Chakraborty}, {and} \bibinfo{person}{Pramod Bhatotia}.} \bibinfo{year}{2022}\natexlab{}.
\newblock \showarticletitle{Risotto: a dynamic binary translator for weak memory model architectures}. In \bibinfo{booktitle}{\emph{Proceedings of the 28th ACM International Conference on Architectural Support for Programming Languages and Operating Systems, Volume 1}}. \bibinfo{pages}{107--122}.
\newblock


\bibitem[Gu and Dao(2023)]%
        {gu2023mamba}
\bibfield{author}{\bibinfo{person}{Albert Gu} {and} \bibinfo{person}{Tri Dao}.} \bibinfo{year}{2023}\natexlab{}.
\newblock \showarticletitle{Mamba: Linear-time sequence modeling with selective state spaces}.
\newblock \bibinfo{journal}{\emph{arXiv preprint arXiv:2312.00752}} (\bibinfo{year}{2023}).
\newblock


\bibitem[Guo et~al\mbox{.}(2024)]%
        {guo2024deepseek}
\bibfield{author}{\bibinfo{person}{Daya Guo}, \bibinfo{person}{Qihao Zhu}, \bibinfo{person}{Dejian Yang}, \bibinfo{person}{Zhenda Xie}, \bibinfo{person}{Kai Dong}, \bibinfo{person}{Wentao Zhang}, \bibinfo{person}{Guanting Chen}, \bibinfo{person}{Xiao Bi}, \bibinfo{person}{Yu Wu}, \bibinfo{person}{YK Li}, {et~al\mbox{.}}} \bibinfo{year}{2024}\natexlab{}.
\newblock \showarticletitle{DeepSeek-Coder: When the Large Language Model Meets Programming--The Rise of Code Intelligence}.
\newblock \bibinfo{journal}{\emph{arXiv preprint arXiv:2401.14196}} (\bibinfo{year}{2024}).
\newblock


\bibitem[Hanif and Maffeis(2022)]%
        {hanif2022vulberta}
\bibfield{author}{\bibinfo{person}{Hazim Hanif} {and} \bibinfo{person}{Sergio Maffeis}.} \bibinfo{year}{2022}\natexlab{}.
\newblock \showarticletitle{Vulberta: Simplified source code pre-training for vulnerability detection}. In \bibinfo{booktitle}{\emph{2022 International joint conference on neural networks (IJCNN)}}. IEEE, \bibinfo{pages}{1--8}.
\newblock


\bibitem[He et~al\mbox{.}(2024)]%
        {he2024code}
\bibfield{author}{\bibinfo{person}{Haojie He}, \bibinfo{person}{Xingwei Lin}, \bibinfo{person}{Ziang Weng}, \bibinfo{person}{Ruijie Zhao}, \bibinfo{person}{Shuitao Gan}, \bibinfo{person}{Libo Chen}, \bibinfo{person}{Yuede Ji}, \bibinfo{person}{Jiashui Wang}, {and} \bibinfo{person}{Zhi Xue}.} \bibinfo{year}{2024}\natexlab{}.
\newblock \showarticletitle{Code is not natural language: Unlock the power of semantics-oriented graph representation for binary code similarity detection}. In \bibinfo{booktitle}{\emph{33rd USENIX Security Symposium (USENIX Security 24), PHILADELPHIA, PA}}.
\newblock


\bibitem[He et~al\mbox{.}(2018)]%
        {he2018debin}
\bibfield{author}{\bibinfo{person}{Jingxuan He}, \bibinfo{person}{Pesho Ivanov}, \bibinfo{person}{Petar Tsankov}, \bibinfo{person}{Veselin Raychev}, {and} \bibinfo{person}{Martin Vechev}.} \bibinfo{year}{2018}\natexlab{}.
\newblock \showarticletitle{Debin: Predicting debug information in stripped binaries}. In \bibinfo{booktitle}{\emph{Proceedings of the 2018 ACM SIGSAC Conference on Computer and Communications Security}}. \bibinfo{pages}{1667--1680}.
\newblock


\bibitem[He and Vechev(2023)]%
        {he2023large}
\bibfield{author}{\bibinfo{person}{Jingxuan He} {and} \bibinfo{person}{Martin Vechev}.} \bibinfo{year}{2023}\natexlab{}.
\newblock \showarticletitle{Large language models for code: Security hardening and adversarial testing}. In \bibinfo{booktitle}{\emph{Proceedings of the 2023 ACM SIGSAC Conference on Computer and Communications Security}}. \bibinfo{pages}{1865--1879}.
\newblock


\bibitem[Hochreiter(1997)]%
        {hochreiter1997long}
\bibfield{author}{\bibinfo{person}{S Hochreiter}.} \bibinfo{year}{1997}\natexlab{}.
\newblock \showarticletitle{Long Short-term Memory}.
\newblock \bibinfo{journal}{\emph{Neural Computation MIT-Press}} (\bibinfo{year}{1997}).
\newblock


\bibitem[Hu et~al\mbox{.}(2024)]%
        {hu2024degpt}
\bibfield{author}{\bibinfo{person}{Peiwei Hu}, \bibinfo{person}{Ruigang Liang}, {and} \bibinfo{person}{Kai Chen}.} \bibinfo{year}{2024}\natexlab{}.
\newblock \showarticletitle{DeGPT: Optimizing Decompiler Output with LLM}. In \bibinfo{booktitle}{\emph{Proceedings 2024 Network and Distributed System Security Symposium (2024). https://api. semanticscholar. org/CorpusID}}, Vol.~\bibinfo{volume}{267622140}.
\newblock


\bibitem[Jiang et~al\mbox{.}(2023)]%
        {jiang2023impact}
\bibfield{author}{\bibinfo{person}{Nan Jiang}, \bibinfo{person}{Kevin Liu}, \bibinfo{person}{Thibaud Lutellier}, {and} \bibinfo{person}{Lin Tan}.} \bibinfo{year}{2023}\natexlab{}.
\newblock \showarticletitle{Impact of code language models on automated program repair}. In \bibinfo{booktitle}{\emph{2023 IEEE/ACM 45th International Conference on Software Engineering (ICSE)}}. IEEE, \bibinfo{pages}{1430--1442}.
\newblock


\bibitem[Kang et~al\mbox{.}(2023)]%
        {kang2023large}
\bibfield{author}{\bibinfo{person}{Sungmin Kang}, \bibinfo{person}{Juyeon Yoon}, {and} \bibinfo{person}{Shin Yoo}.} \bibinfo{year}{2023}\natexlab{}.
\newblock \showarticletitle{Large language models are few-shot testers: Exploring llm-based general bug reproduction}. In \bibinfo{booktitle}{\emph{2023 IEEE/ACM 45th International Conference on Software Engineering (ICSE)}}. IEEE, \bibinfo{pages}{2312--2323}.
\newblock


\bibitem[Kim et~al\mbox{.}(2022)]%
        {kim2022revisiting}
\bibfield{author}{\bibinfo{person}{Dongkwan Kim}, \bibinfo{person}{Eunsoo Kim}, \bibinfo{person}{Sang~Kil Cha}, \bibinfo{person}{Sooel Son}, {and} \bibinfo{person}{Yongdae Kim}.} \bibinfo{year}{2022}\natexlab{}.
\newblock \showarticletitle{Revisiting binary code similarity analysis using interpretable feature engineering and lessons learned}.
\newblock \bibinfo{journal}{\emph{IEEE Transactions on Software Engineering}} \bibinfo{volume}{49}, \bibinfo{number}{4} (\bibinfo{year}{2022}), \bibinfo{pages}{1661--1682}.
\newblock


\bibitem[Kim et~al\mbox{.}(2023)]%
        {kim2023funprobe}
\bibfield{author}{\bibinfo{person}{Soomin Kim}, \bibinfo{person}{Hyungseok Kim}, {and} \bibinfo{person}{Sang~Kil Cha}.} \bibinfo{year}{2023}\natexlab{}.
\newblock \showarticletitle{Funprobe: Probing functions from binary code through probabilistic analysis}. In \bibinfo{booktitle}{\emph{Proceedings of the 31st ACM Joint European Software Engineering Conference and Symposium on the Foundations of Software Engineering}}. \bibinfo{pages}{1419--1430}.
\newblock


\bibitem[Lee et~al\mbox{.}(2011)]%
        {lee2011tie}
\bibfield{author}{\bibinfo{person}{JongHyup Lee}, \bibinfo{person}{Thanassis Avgerinos}, {and} \bibinfo{person}{David Brumley}.} \bibinfo{year}{2011}\natexlab{}.
\newblock \showarticletitle{TIE: Principled reverse engineering of types in binary programs}.
\newblock  (\bibinfo{year}{2011}).
\newblock


\bibitem[Li et~al\mbox{.}(2023b)]%
        {li2023vulanalyzer}
\bibfield{author}{\bibinfo{person}{Litao Li}, \bibinfo{person}{Steven~HH Ding}, \bibinfo{person}{Yuan Tian}, \bibinfo{person}{Benjamin~CM Fung}, \bibinfo{person}{Philippe Charland}, \bibinfo{person}{Weihan Ou}, \bibinfo{person}{Leo Song}, {and} \bibinfo{person}{Congwei Chen}.} \bibinfo{year}{2023}\natexlab{b}.
\newblock \showarticletitle{VulANalyzeR: Explainable binary vulnerability detection with multi-task learning and attentional graph convolution}.
\newblock \bibinfo{journal}{\emph{ACM Transactions on Privacy and Security}} \bibinfo{volume}{26}, \bibinfo{number}{3} (\bibinfo{year}{2023}), \bibinfo{pages}{1--25}.
\newblock


\bibitem[Li et~al\mbox{.}(2023a)]%
        {li2023starcoder}
\bibfield{author}{\bibinfo{person}{Raymond Li}, \bibinfo{person}{Loubna~Ben Allal}, \bibinfo{person}{Yangtian Zi}, \bibinfo{person}{Niklas Muennighoff}, \bibinfo{person}{Denis Kocetkov}, \bibinfo{person}{Chenghao Mou}, \bibinfo{person}{Marc Marone}, \bibinfo{person}{Christopher Akiki}, \bibinfo{person}{Jia Li}, \bibinfo{person}{Jenny Chim}, {et~al\mbox{.}}} \bibinfo{year}{2023}\natexlab{a}.
\newblock \showarticletitle{Starcoder: may the source be with you!}
\newblock \bibinfo{journal}{\emph{arXiv preprint arXiv:2305.06161}} (\bibinfo{year}{2023}).
\newblock


\bibitem[Li et~al\mbox{.}(2023c)]%
        {li2023cctest}
\bibfield{author}{\bibinfo{person}{Zongjie Li}, \bibinfo{person}{Chaozheng Wang}, \bibinfo{person}{Zhibo Liu}, \bibinfo{person}{Haoxuan Wang}, \bibinfo{person}{Dong Chen}, \bibinfo{person}{Shuai Wang}, {and} \bibinfo{person}{Cuiyun Gao}.} \bibinfo{year}{2023}\natexlab{c}.
\newblock \showarticletitle{Cctest: Testing and repairing code completion systems}. In \bibinfo{booktitle}{\emph{2023 IEEE/ACM 45th International Conference on Software Engineering (ICSE)}}. IEEE, \bibinfo{pages}{1238--1250}.
\newblock


\bibitem[Liu et~al\mbox{.}(2023)]%
        {liu2023harnessing}
\bibfield{author}{\bibinfo{person}{Puzhuo Liu}, \bibinfo{person}{Chengnian Sun}, \bibinfo{person}{Yaowen Zheng}, \bibinfo{person}{Xuan Feng}, \bibinfo{person}{Chuan Qin}, \bibinfo{person}{Yuncheng Wang}, \bibinfo{person}{Zhi Li}, {and} \bibinfo{person}{Limin Sun}.} \bibinfo{year}{2023}\natexlab{}.
\newblock \showarticletitle{Harnessing the power of llm to support binary taint analysis}.
\newblock \bibinfo{journal}{\emph{arXiv preprint arXiv:2310.08275}} (\bibinfo{year}{2023}).
\newblock


\bibitem[Liu(2019)]%
        {liu2019roberta}
\bibfield{author}{\bibinfo{person}{Yinhan Liu}.} \bibinfo{year}{2019}\natexlab{}.
\newblock \showarticletitle{Roberta: A robustly optimized bert pretraining approach}.
\newblock \bibinfo{journal}{\emph{arXiv preprint arXiv:1907.11692}}  \bibinfo{volume}{364} (\bibinfo{year}{2019}).
\newblock


\bibitem[Liu et~al\mbox{.}(2024)]%
        {liu2024pre}
\bibfield{author}{\bibinfo{person}{Zhongxin Liu}, \bibinfo{person}{Zhijie Tang}, \bibinfo{person}{Junwei Zhang}, \bibinfo{person}{Xin Xia}, {and} \bibinfo{person}{Xiaohu Yang}.} \bibinfo{year}{2024}\natexlab{}.
\newblock \showarticletitle{Pre-training by Predicting Program Dependencies for Vulnerability Analysis Tasks}. In \bibinfo{booktitle}{\emph{Proceedings of the IEEE/ACM 46th International Conference on Software Engineering}}. \bibinfo{pages}{1--13}.
\newblock


\bibitem[Luo et~al\mbox{.}(2023)]%
        {luo2023vulhawk}
\bibfield{author}{\bibinfo{person}{Zhenhao Luo}, \bibinfo{person}{Pengfei Wang}, \bibinfo{person}{Baosheng Wang}, \bibinfo{person}{Yong Tang}, \bibinfo{person}{Wei Xie}, \bibinfo{person}{Xu Zhou}, \bibinfo{person}{Danjun Liu}, {and} \bibinfo{person}{Kai Lu}.} \bibinfo{year}{2023}\natexlab{}.
\newblock \showarticletitle{VulHawk: Cross-architecture Vulnerability Detection with Entropy-based Binary Code Search.}. In \bibinfo{booktitle}{\emph{NDSS}}.
\newblock


\bibitem[Madaan et~al\mbox{.}(2024)]%
        {madaan2024self}
\bibfield{author}{\bibinfo{person}{Aman Madaan}, \bibinfo{person}{Niket Tandon}, \bibinfo{person}{Prakhar Gupta}, \bibinfo{person}{Skyler Hallinan}, \bibinfo{person}{Luyu Gao}, \bibinfo{person}{Sarah Wiegreffe}, \bibinfo{person}{Uri Alon}, \bibinfo{person}{Nouha Dziri}, \bibinfo{person}{Shrimai Prabhumoye}, \bibinfo{person}{Yiming Yang}, {et~al\mbox{.}}} \bibinfo{year}{2024}\natexlab{}.
\newblock \showarticletitle{Self-refine: Iterative refinement with self-feedback}.
\newblock \bibinfo{journal}{\emph{Advances in Neural Information Processing Systems}}  \bibinfo{volume}{36} (\bibinfo{year}{2024}).
\newblock


\bibitem[Maheswaranathan et~al\mbox{.}(2021)]%
        {maheswaranathan2021reverse}
\bibfield{author}{\bibinfo{person}{Niru Maheswaranathan}, \bibinfo{person}{David Sussillo}, \bibinfo{person}{Luke Metz}, \bibinfo{person}{Ruoxi Sun}, {and} \bibinfo{person}{Jascha Sohl-Dickstein}.} \bibinfo{year}{2021}\natexlab{}.
\newblock \showarticletitle{Reverse engineering learned optimizers reveals known and novel mechanisms}.
\newblock \bibinfo{journal}{\emph{Advances in Neural Information Processing Systems}}  \bibinfo{volume}{34} (\bibinfo{year}{2021}), \bibinfo{pages}{19910--19922}.
\newblock


\bibitem[Mirsky et~al\mbox{.}(2023)]%
        {mirsky2023vulchecker}
\bibfield{author}{\bibinfo{person}{Yisroel Mirsky}, \bibinfo{person}{George Macon}, \bibinfo{person}{Michael Brown}, \bibinfo{person}{Carter Yagemann}, \bibinfo{person}{Matthew Pruett}, \bibinfo{person}{Evan Downing}, \bibinfo{person}{Sukarno Mertoguno}, {and} \bibinfo{person}{Wenke Lee}.} \bibinfo{year}{2023}\natexlab{}.
\newblock \showarticletitle{$\{$VulChecker$\}$: Graph-based Vulnerability Localization in Source Code}. In \bibinfo{booktitle}{\emph{32nd USENIX Security Symposium (USENIX Security 23)}}. \bibinfo{pages}{6557--6574}.
\newblock


\bibitem[{MITRE}(2024)]%
        {cve}
\bibfield{author}{\bibinfo{person}{{MITRE}}.} \bibinfo{year}{2024}\natexlab{}.
\newblock \bibinfo{booktitle}{\emph{Common Vulnerabilities and Exposures}}.
\newblock
\urldef\tempurl%
\url{https://cve.mitre.org/}
\showURL{%
\tempurl}


\bibitem[MITRE(2024a)]%
        {cwe}
\bibfield{author}{\bibinfo{person}{MITRE}.} \bibinfo{year}{2024}\natexlab{a}.
\newblock \bibinfo{booktitle}{\emph{Common Weakness Enumeration}}.
\newblock
\urldef\tempurl%
\url{https://cwe.mitre.org/}
\showURL{%
\tempurl}


\bibitem[MITRE(2024b)]%
        {juliet}
\bibfield{author}{\bibinfo{person}{MITRE}.} \bibinfo{year}{2024}\natexlab{b}.
\newblock \bibinfo{booktitle}{\emph{Juliet C/C++ 1.3 v1.3}}.
\newblock
\urldef\tempurl%
\url{https://samate.nist.gov/SARD/test-suites/112}
\showURL{%
\tempurl}


\bibitem[{MITRE}(2024a)]%
        {cve-2023-3609}
\bibfield{author}{\bibinfo{person}{{MITRE}}.} \bibinfo{year}{2024}\natexlab{a}.
\newblock \bibinfo{booktitle}{\emph{The webiste of cve-2023-3699}}.
\newblock
\urldef\tempurl%
\url{https://cve.mitre.org/cgi-bin/cvename.cgi?name=CVE-2023-3609}
\showURL{%
\tempurl}


\bibitem[{MITRE}(2024b)]%
        {cve-2023-30772}
\bibfield{author}{\bibinfo{person}{{MITRE}}.} \bibinfo{year}{2024}\natexlab{b}.
\newblock \bibinfo{booktitle}{\emph{The website of cve-2023-30772}}.
\newblock
\urldef\tempurl%
\url{https://cve.mitre.org/cgi-bin/cvename.cgi?name=CVE-2023-30772}
\showURL{%
\tempurl}


\bibitem[MITRE(2024)]%
        {cwe416}
\bibfield{author}{\bibinfo{person}{MITRE}.} \bibinfo{year}{2024}\natexlab{}.
\newblock \bibinfo{booktitle}{\emph{The website of cwe-416}}.
\newblock
\urldef\tempurl%
\url{https://cwe.mitre.org/data/definitions/416.html}
\showURL{%
\tempurl}


\bibitem[Muennighoff et~al\mbox{.}(2023)]%
        {muennighoff2023octopack}
\bibfield{author}{\bibinfo{person}{Niklas Muennighoff}, \bibinfo{person}{Qian Liu}, \bibinfo{person}{Armel Zebaze}, \bibinfo{person}{Qinkai Zheng}, \bibinfo{person}{Binyuan Hui}, \bibinfo{person}{Terry~Yue Zhuo}, \bibinfo{person}{Swayam Singh}, \bibinfo{person}{Xiangru Tang}, \bibinfo{person}{Leandro Von~Werra}, {and} \bibinfo{person}{Shayne Longpre}.} \bibinfo{year}{2023}\natexlab{}.
\newblock \showarticletitle{Octopack: Instruction tuning code large language models}.
\newblock \bibinfo{journal}{\emph{arXiv preprint arXiv:2308.07124}} (\bibinfo{year}{2023}).
\newblock


\bibitem[Nam et~al\mbox{.}(2024)]%
        {nam2024using}
\bibfield{author}{\bibinfo{person}{Daye Nam}, \bibinfo{person}{Andrew Macvean}, \bibinfo{person}{Vincent Hellendoorn}, \bibinfo{person}{Bogdan Vasilescu}, {and} \bibinfo{person}{Brad Myers}.} \bibinfo{year}{2024}\natexlab{}.
\newblock \showarticletitle{Using an llm to help with code understanding}. In \bibinfo{booktitle}{\emph{Proceedings of the IEEE/ACM 46th International Conference on Software Engineering}}. \bibinfo{pages}{1--13}.
\newblock


\bibitem[Neelakantan et~al\mbox{.}(2022)]%
        {neelakantan2022text}
\bibfield{author}{\bibinfo{person}{Arvind Neelakantan}, \bibinfo{person}{Tao Xu}, \bibinfo{person}{Raul Puri}, \bibinfo{person}{Alec Radford}, \bibinfo{person}{Jesse~Michael Han}, \bibinfo{person}{Jerry Tworek}, \bibinfo{person}{Qiming Yuan}, \bibinfo{person}{Nikolas Tezak}, \bibinfo{person}{Jong~Wook Kim}, \bibinfo{person}{Chris Hallacy}, {et~al\mbox{.}}} \bibinfo{year}{2022}\natexlab{}.
\newblock \showarticletitle{Text and code embeddings by contrastive pre-training}.
\newblock \bibinfo{journal}{\emph{arXiv preprint arXiv:2201.10005}} (\bibinfo{year}{2022}).
\newblock


\bibitem[Nijkamp et~al\mbox{.}(2022)]%
        {nijkamp2022codegen}
\bibfield{author}{\bibinfo{person}{Erik Nijkamp}, \bibinfo{person}{Bo Pang}, \bibinfo{person}{Hiroaki Hayashi}, \bibinfo{person}{Lifu Tu}, \bibinfo{person}{Huan Wang}, \bibinfo{person}{Yingbo Zhou}, \bibinfo{person}{Silvio Savarese}, {and} \bibinfo{person}{Caiming Xiong}.} \bibinfo{year}{2022}\natexlab{}.
\newblock \showarticletitle{Codegen: An open large language model for code with multi-turn program synthesis}.
\newblock \bibinfo{journal}{\emph{arXiv preprint arXiv:2203.13474}} (\bibinfo{year}{2022}).
\newblock


\bibitem[Nitin et~al\mbox{.}(2021)]%
        {nitin2021direct}
\bibfield{author}{\bibinfo{person}{Vikram Nitin}, \bibinfo{person}{Anthony Saieva}, \bibinfo{person}{Baishakhi Ray}, {and} \bibinfo{person}{Gail Kaiser}.} \bibinfo{year}{2021}\natexlab{}.
\newblock \showarticletitle{Direct: A transformer-based model for decompiled identifier renaming}. In \bibinfo{booktitle}{\emph{Proceedings of the 1st Workshop on Natural Language Processing for Programming (NLP4Prog 2021)}}. \bibinfo{pages}{48--57}.
\newblock


\bibitem[Noonan et~al\mbox{.}(2016)]%
        {noonan2016polymorphic}
\bibfield{author}{\bibinfo{person}{Matt Noonan}, \bibinfo{person}{Alexey Loginov}, {and} \bibinfo{person}{David Cok}.} \bibinfo{year}{2016}\natexlab{}.
\newblock \showarticletitle{Polymorphic type inference for machine code}. In \bibinfo{booktitle}{\emph{Proceedings of the 37th ACM SIGPLAN Conference on Programming Language Design and Implementation}}. \bibinfo{pages}{27--41}.
\newblock


\bibitem[Olausson et~al\mbox{.}(2023)]%
        {olausson2023demystifying}
\bibfield{author}{\bibinfo{person}{Theo~X Olausson}, \bibinfo{person}{Jeevana~Priya Inala}, \bibinfo{person}{Chenglong Wang}, \bibinfo{person}{Jianfeng Gao}, {and} \bibinfo{person}{Armando Solar-Lezama}.} \bibinfo{year}{2023}\natexlab{}.
\newblock \showarticletitle{Demystifying gpt self-repair for code generation}.
\newblock \bibinfo{journal}{\emph{arXiv preprint arXiv:2306.09896}} (\bibinfo{year}{2023}).
\newblock


\bibitem[OpenAI(2024a)]%
        {openai2024chatgpt}
\bibfield{author}{\bibinfo{person}{OpenAI}.} \bibinfo{year}{2024}\natexlab{a}.
\newblock \bibinfo{booktitle}{\emph{ChatGPT}}.
\newblock
\urldef\tempurl%
\url{https://openai.com/blog/chatgpt}
\showURL{%
\tempurl}


\bibitem[OpenAI(2024b)]%
        {gpt4o}
\bibfield{author}{\bibinfo{person}{OpenAI}.} \bibinfo{year}{2024}\natexlab{b}.
\newblock \bibinfo{booktitle}{\emph{GPT-4o}}.
\newblock
\urldef\tempurl%
\url{https://openai.com/index/hello-gpt-4o/}
\showURL{%
\tempurl}


\bibitem[Pang et~al\mbox{.}(2021)]%
        {pang2021sok}
\bibfield{author}{\bibinfo{person}{Chengbin Pang}, \bibinfo{person}{Ruotong Yu}, \bibinfo{person}{Yaohui Chen}, \bibinfo{person}{Eric Koskinen}, \bibinfo{person}{Georgios Portokalidis}, \bibinfo{person}{Bing Mao}, {and} \bibinfo{person}{Jun Xu}.} \bibinfo{year}{2021}\natexlab{}.
\newblock \showarticletitle{Sok: All you ever wanted to know about x86/x64 binary disassembly but were afraid to ask}. In \bibinfo{booktitle}{\emph{2021 IEEE symposium on security and privacy (SP)}}. IEEE, \bibinfo{pages}{833--851}.
\newblock


\bibitem[Pearce et~al\mbox{.}(2023)]%
        {pearce2023examining}
\bibfield{author}{\bibinfo{person}{Hammond Pearce}, \bibinfo{person}{Benjamin Tan}, \bibinfo{person}{Baleegh Ahmad}, \bibinfo{person}{Ramesh Karri}, {and} \bibinfo{person}{Brendan Dolan-Gavitt}.} \bibinfo{year}{2023}\natexlab{}.
\newblock \showarticletitle{Examining zero-shot vulnerability repair with large language models}. In \bibinfo{booktitle}{\emph{2023 IEEE Symposium on Security and Privacy (SP)}}. IEEE, \bibinfo{pages}{2339--2356}.
\newblock


\bibitem[Pei et~al\mbox{.}(2022a)]%
        {pei2022neudep}
\bibfield{author}{\bibinfo{person}{Kexin Pei}, \bibinfo{person}{Dongdong She}, \bibinfo{person}{Michael Wang}, \bibinfo{person}{Scott Geng}, \bibinfo{person}{Zhou Xuan}, \bibinfo{person}{Yaniv David}, \bibinfo{person}{Junfeng Yang}, \bibinfo{person}{Suman Jana}, {and} \bibinfo{person}{Baishakhi Ray}.} \bibinfo{year}{2022}\natexlab{a}.
\newblock \showarticletitle{NeuDep: neural binary memory dependence analysis}. In \bibinfo{booktitle}{\emph{Proceedings of the 30th ACM Joint European Software Engineering Conference and Symposium on the Foundations of Software Engineering}}. \bibinfo{pages}{747--759}.
\newblock


\bibitem[Pei et~al\mbox{.}(2022b)]%
        {pei2022learning}
\bibfield{author}{\bibinfo{person}{Kexin Pei}, \bibinfo{person}{Zhou Xuan}, \bibinfo{person}{Junfeng Yang}, \bibinfo{person}{Suman Jana}, {and} \bibinfo{person}{Baishakhi Ray}.} \bibinfo{year}{2022}\natexlab{b}.
\newblock \showarticletitle{Learning approximate execution semantics from traces for binary function similarity}.
\newblock \bibinfo{journal}{\emph{IEEE Transactions on Software Engineering}} \bibinfo{volume}{49}, \bibinfo{number}{4} (\bibinfo{year}{2022}), \bibinfo{pages}{2776--2790}.
\newblock


\bibitem[Peng et~al\mbox{.}(2023)]%
        {peng2023ptlvd}
\bibfield{author}{\bibinfo{person}{Tao Peng}, \bibinfo{person}{Shixu Chen}, \bibinfo{person}{Fei Zhu}, \bibinfo{person}{Junwei Tang}, \bibinfo{person}{Junping Liu}, {and} \bibinfo{person}{Xinrong Hu}.} \bibinfo{year}{2023}\natexlab{}.
\newblock \showarticletitle{PTLVD: Program Slicing and Transformer-based Line-level Vulnerability Detection System}. In \bibinfo{booktitle}{\emph{2023 IEEE 23rd International Working Conference on Source Code Analysis and Manipulation (SCAM)}}. IEEE, \bibinfo{pages}{162--173}.
\newblock


\bibitem[Peng et~al\mbox{.}(2024)]%
        {peng2024domain}
\bibfield{author}{\bibinfo{person}{Yun Peng}, \bibinfo{person}{Shuzheng Gao}, \bibinfo{person}{Cuiyun Gao}, \bibinfo{person}{Yintong Huo}, {and} \bibinfo{person}{Michael Lyu}.} \bibinfo{year}{2024}\natexlab{}.
\newblock \showarticletitle{Domain knowledge matters: Improving prompts with fix templates for repairing python type errors}. In \bibinfo{booktitle}{\emph{Proceedings of the 46th IEEE/ACM International Conference on Software Engineering}}. \bibinfo{pages}{1--13}.
\newblock


\bibitem[Radford(2018)]%
        {radford2018improving}
\bibfield{author}{\bibinfo{person}{Alec Radford}.} \bibinfo{year}{2018}\natexlab{}.
\newblock \showarticletitle{Improving language understanding by generative pre-training}.
\newblock  (\bibinfo{year}{2018}).
\newblock


\bibitem[Roziere et~al\mbox{.}(2023)]%
        {roziere2023code}
\bibfield{author}{\bibinfo{person}{Baptiste Roziere}, \bibinfo{person}{Jonas Gehring}, \bibinfo{person}{Fabian Gloeckle}, \bibinfo{person}{Sten Sootla}, \bibinfo{person}{Itai Gat}, \bibinfo{person}{Xiaoqing~Ellen Tan}, \bibinfo{person}{Yossi Adi}, \bibinfo{person}{Jingyu Liu}, \bibinfo{person}{Romain Sauvestre}, \bibinfo{person}{Tal Remez}, {et~al\mbox{.}}} \bibinfo{year}{2023}\natexlab{}.
\newblock \showarticletitle{Code llama: Open foundation models for code}.
\newblock \bibinfo{journal}{\emph{arXiv preprint arXiv:2308.12950}} (\bibinfo{year}{2023}).
\newblock


\bibitem[Scheurer et~al\mbox{.}(2023)]%
        {scheurer2023training}
\bibfield{author}{\bibinfo{person}{J{\'e}r{\'e}my Scheurer}, \bibinfo{person}{Jon~Ander Campos}, \bibinfo{person}{Tomasz Korbak}, \bibinfo{person}{Jun~Shern Chan}, \bibinfo{person}{Angelica Chen}, \bibinfo{person}{Kyunghyun Cho}, {and} \bibinfo{person}{Ethan Perez}.} \bibinfo{year}{2023}\natexlab{}.
\newblock \showarticletitle{Training language models with language feedback at scale}.
\newblock \bibinfo{journal}{\emph{arXiv preprint arXiv:2303.16755}} (\bibinfo{year}{2023}).
\newblock


\bibitem[Shoshitaishvili et~al\mbox{.}(2016)]%
        {shoshitaishvili2016sok}
\bibfield{author}{\bibinfo{person}{Yan Shoshitaishvili}, \bibinfo{person}{Ruoyu Wang}, \bibinfo{person}{Christopher Salls}, \bibinfo{person}{Nick Stephens}, \bibinfo{person}{Mario Polino}, \bibinfo{person}{Andrew Dutcher}, \bibinfo{person}{John Grosen}, \bibinfo{person}{Siji Feng}, \bibinfo{person}{Christophe Hauser}, \bibinfo{person}{Christopher Kruegel}, {et~al\mbox{.}}} \bibinfo{year}{2016}\natexlab{}.
\newblock \showarticletitle{Sok:(state of) the art of war: Offensive techniques in binary analysis}. In \bibinfo{booktitle}{\emph{2016 IEEE symposium on security and privacy (SP)}}. IEEE, \bibinfo{pages}{138--157}.
\newblock


\bibitem[Steenhoek et~al\mbox{.}(2024)]%
        {steenhoek2024dataflow}
\bibfield{author}{\bibinfo{person}{Benjamin Steenhoek}, \bibinfo{person}{Hongyang Gao}, {and} \bibinfo{person}{Wei Le}.} \bibinfo{year}{2024}\natexlab{}.
\newblock \showarticletitle{Dataflow analysis-inspired deep learning for efficient vulnerability detection}. In \bibinfo{booktitle}{\emph{Proceedings of the 46th IEEE/ACM International Conference on Software Engineering}}. \bibinfo{pages}{1--13}.
\newblock


\bibitem[Steenhoek et~al\mbox{.}(2023)]%
        {steenhoek2023empirical}
\bibfield{author}{\bibinfo{person}{Benjamin Steenhoek}, \bibinfo{person}{Md~Mahbubur Rahman}, \bibinfo{person}{Richard Jiles}, {and} \bibinfo{person}{Wei Le}.} \bibinfo{year}{2023}\natexlab{}.
\newblock \showarticletitle{An empirical study of deep learning models for vulnerability detection}. In \bibinfo{booktitle}{\emph{2023 IEEE/ACM 45th International Conference on Software Engineering (ICSE)}}. IEEE, \bibinfo{pages}{2237--2248}.
\newblock


\bibitem[Su et~al\mbox{.}(2024a)]%
        {su2024source}
\bibfield{author}{\bibinfo{person}{Zian Su}, \bibinfo{person}{Xiangzhe Xu}, \bibinfo{person}{Ziyang Huang}, \bibinfo{person}{Kaiyuan Zhang}, {and} \bibinfo{person}{Xiangyu Zhang}.} \bibinfo{year}{2024}\natexlab{a}.
\newblock \showarticletitle{Source Code Foundation Models are Transferable Binary Analysis Knowledge Bases}.
\newblock \bibinfo{journal}{\emph{arXiv preprint arXiv:2405.19581}} (\bibinfo{year}{2024}).
\newblock


\bibitem[Su et~al\mbox{.}(2024b)]%
        {su2024codeart}
\bibfield{author}{\bibinfo{person}{Zian Su}, \bibinfo{person}{Xiangzhe Xu}, \bibinfo{person}{Ziyang Huang}, \bibinfo{person}{Zhuo Zhang}, \bibinfo{person}{Yapeng Ye}, \bibinfo{person}{Jianjun Huang}, {and} \bibinfo{person}{Xiangyu Zhang}.} \bibinfo{year}{2024}\natexlab{b}.
\newblock \showarticletitle{Codeart: Better code models by attention regularization when symbols are lacking}.
\newblock \bibinfo{journal}{\emph{Proceedings of the ACM on Software Engineering}} \bibinfo{volume}{1}, \bibinfo{number}{FSE} (\bibinfo{year}{2024}), \bibinfo{pages}{562--585}.
\newblock


\bibitem[Talebirad and Nadiri(2023)]%
        {talebirad2023multi}
\bibfield{author}{\bibinfo{person}{Yashar Talebirad} {and} \bibinfo{person}{Amirhossein Nadiri}.} \bibinfo{year}{2023}\natexlab{}.
\newblock \showarticletitle{Multi-agent collaboration: Harnessing the power of intelligent llm agents}.
\newblock \bibinfo{journal}{\emph{arXiv preprint arXiv:2306.03314}} (\bibinfo{year}{2023}).
\newblock


\bibitem[Tan et~al\mbox{.}(2024)]%
        {tan2024llm4decompile}
\bibfield{author}{\bibinfo{person}{Hanzhuo Tan}, \bibinfo{person}{Qi Luo}, \bibinfo{person}{Jing Li}, {and} \bibinfo{person}{Yuqun Zhang}.} \bibinfo{year}{2024}\natexlab{}.
\newblock \showarticletitle{LLM4Decompile: Decompiling Binary Code with Large Language Models}.
\newblock \bibinfo{journal}{\emph{arXiv preprint arXiv:2403.05286}} (\bibinfo{year}{2024}).
\newblock


\bibitem[Tang et~al\mbox{.}(2023)]%
        {tang2023csgvd}
\bibfield{author}{\bibinfo{person}{Wei Tang}, \bibinfo{person}{Mingwei Tang}, \bibinfo{person}{Minchao Ban}, \bibinfo{person}{Ziguo Zhao}, {and} \bibinfo{person}{Mingjun Feng}.} \bibinfo{year}{2023}\natexlab{}.
\newblock \showarticletitle{CSGVD: A deep learning approach combining sequence and graph embedding for source code vulnerability detection}.
\newblock \bibinfo{journal}{\emph{Journal of Systems and Software}}  \bibinfo{volume}{199} (\bibinfo{year}{2023}), \bibinfo{pages}{111623}.
\newblock


\bibitem[Team et~al\mbox{.}(2023)]%
        {team2023gemini}
\bibfield{author}{\bibinfo{person}{Gemini Team}, \bibinfo{person}{Rohan Anil}, \bibinfo{person}{Sebastian Borgeaud}, \bibinfo{person}{Jean-Baptiste Alayrac}, \bibinfo{person}{Jiahui Yu}, \bibinfo{person}{Radu Soricut}, \bibinfo{person}{Johan Schalkwyk}, \bibinfo{person}{Andrew~M Dai}, \bibinfo{person}{Anja Hauth}, \bibinfo{person}{Katie Millican}, {et~al\mbox{.}}} \bibinfo{year}{2023}\natexlab{}.
\newblock \showarticletitle{Gemini: a family of highly capable multimodal models}.
\newblock \bibinfo{journal}{\emph{arXiv preprint arXiv:2312.11805}} (\bibinfo{year}{2023}).
\newblock


\bibitem[Touvron et~al\mbox{.}(2023)]%
        {touvron2023llama}
\bibfield{author}{\bibinfo{person}{Hugo Touvron}, \bibinfo{person}{Louis Martin}, \bibinfo{person}{Kevin Stone}, \bibinfo{person}{Peter Albert}, \bibinfo{person}{Amjad Almahairi}, \bibinfo{person}{Yasmine Babaei}, \bibinfo{person}{Nikolay Bashlykov}, \bibinfo{person}{Soumya Batra}, \bibinfo{person}{Prajjwal Bhargava}, \bibinfo{person}{Shruti Bhosale}, {et~al\mbox{.}}} \bibinfo{year}{2023}\natexlab{}.
\newblock \showarticletitle{Llama 2: Open foundation and fine-tuned chat models}.
\newblock \bibinfo{journal}{\emph{arXiv preprint arXiv:2307.09288}} (\bibinfo{year}{2023}).
\newblock


\bibitem[Ullah et~al\mbox{.}(2024)]%
        {ullah2024llms}
\bibfield{author}{\bibinfo{person}{Saad Ullah}, \bibinfo{person}{Mingji Han}, \bibinfo{person}{Saurabh Pujar}, \bibinfo{person}{Hammond Pearce}, \bibinfo{person}{Ayse Coskun}, {and} \bibinfo{person}{Gianluca Stringhini}.} \bibinfo{year}{2024}\natexlab{}.
\newblock \showarticletitle{LLMs Cannot Reliably Identify and Reason About Security Vulnerabilities (Yet?): A Comprehensive Evaluation, Framework, and Benchmarks}. In \bibinfo{booktitle}{\emph{IEEE Symposium on Security and Privacy}}.
\newblock


\bibitem[Vadayath et~al\mbox{.}(2022)]%
        {vadayath2022arbiter}
\bibfield{author}{\bibinfo{person}{Jayakrishna Vadayath}, \bibinfo{person}{Moritz Eckert}, \bibinfo{person}{Kyle Zeng}, \bibinfo{person}{Nicolaas Weideman}, \bibinfo{person}{Gokulkrishna~Praveen Menon}, \bibinfo{person}{Yanick Fratantonio}, \bibinfo{person}{Davide Balzarotti}, \bibinfo{person}{Adam Doup{\'e}}, \bibinfo{person}{Tiffany Bao}, \bibinfo{person}{Ruoyu Wang}, {et~al\mbox{.}}} \bibinfo{year}{2022}\natexlab{}.
\newblock \showarticletitle{Arbiter: Bridging the static and dynamic divide in vulnerability discovery on binary programs}. In \bibinfo{booktitle}{\emph{31st USENIX Security Symposium (USENIX Security 22)}}. \bibinfo{pages}{413--430}.
\newblock


\bibitem[Vaswani(2017)]%
        {vaswani2017attention}
\bibfield{author}{\bibinfo{person}{A Vaswani}.} \bibinfo{year}{2017}\natexlab{}.
\newblock \showarticletitle{Attention is all you need}.
\newblock \bibinfo{journal}{\emph{Advances in Neural Information Processing Systems}} (\bibinfo{year}{2017}).
\newblock


\bibitem[Wang and Shoshitaishvili(2017)]%
        {wang2017angr}
\bibfield{author}{\bibinfo{person}{Fish Wang} {and} \bibinfo{person}{Yan Shoshitaishvili}.} \bibinfo{year}{2017}\natexlab{}.
\newblock \showarticletitle{Angr-the next generation of binary analysis}. In \bibinfo{booktitle}{\emph{2017 IEEE Cybersecurity Development (SecDev)}}. IEEE, \bibinfo{pages}{8--9}.
\newblock


\bibitem[Wang et~al\mbox{.}(2022)]%
        {wang2022jtrans}
\bibfield{author}{\bibinfo{person}{Hao Wang}, \bibinfo{person}{Wenjie Qu}, \bibinfo{person}{Gilad Katz}, \bibinfo{person}{Wenyu Zhu}, \bibinfo{person}{Zeyu Gao}, \bibinfo{person}{Han Qiu}, \bibinfo{person}{Jianwei Zhuge}, {and} \bibinfo{person}{Chao Zhang}.} \bibinfo{year}{2022}\natexlab{}.
\newblock \showarticletitle{Jtrans: Jump-aware transformer for binary code similarity detection}. In \bibinfo{booktitle}{\emph{Proceedings of the 31st ACM SIGSOFT International Symposium on Software Testing and Analysis}}. \bibinfo{pages}{1--13}.
\newblock


\bibitem[Wang et~al\mbox{.}(2023)]%
        {wang2023can}
\bibfield{author}{\bibinfo{person}{Junzhe Wang}, \bibinfo{person}{Matthew Sharp}, \bibinfo{person}{Chuxiong Wu}, \bibinfo{person}{Qiang Zeng}, {and} \bibinfo{person}{Lannan Luo}.} \bibinfo{year}{2023}\natexlab{}.
\newblock \showarticletitle{Can a Deep Learning Model for One Architecture Be Used for Others?$\{$Retargeted-Architecture$\}$ Binary Code Analysis}. In \bibinfo{booktitle}{\emph{32nd USENIX Security Symposium (USENIX Security 23)}}. \bibinfo{pages}{7339--7356}.
\newblock


\bibitem[Wang et~al\mbox{.}(2021)]%
        {wang2021codet5}
\bibfield{author}{\bibinfo{person}{Yue Wang}, \bibinfo{person}{Weishi Wang}, \bibinfo{person}{Shafiq Joty}, {and} \bibinfo{person}{Steven~CH Hoi}.} \bibinfo{year}{2021}\natexlab{}.
\newblock \showarticletitle{Codet5: Identifier-aware unified pre-trained encoder-decoder models for code understanding and generation}.
\newblock \bibinfo{journal}{\emph{arXiv preprint arXiv:2109.00859}} (\bibinfo{year}{2021}).
\newblock


\bibitem[Wei et~al\mbox{.}(2022)]%
        {wei2022chain}
\bibfield{author}{\bibinfo{person}{Jason Wei}, \bibinfo{person}{Xuezhi Wang}, \bibinfo{person}{Dale Schuurmans}, \bibinfo{person}{Maarten Bosma}, \bibinfo{person}{Fei Xia}, \bibinfo{person}{Ed Chi}, \bibinfo{person}{Quoc~V Le}, \bibinfo{person}{Denny Zhou}, {et~al\mbox{.}}} \bibinfo{year}{2022}\natexlab{}.
\newblock \showarticletitle{Chain-of-thought prompting elicits reasoning in large language models}.
\newblock \bibinfo{journal}{\emph{Advances in neural information processing systems}}  \bibinfo{volume}{35} (\bibinfo{year}{2022}), \bibinfo{pages}{24824--24837}.
\newblock


\bibitem[Wu et~al\mbox{.}(2023)]%
        {wu2023effective}
\bibfield{author}{\bibinfo{person}{Yi Wu}, \bibinfo{person}{Nan Jiang}, \bibinfo{person}{Hung~Viet Pham}, \bibinfo{person}{Thibaud Lutellier}, \bibinfo{person}{Jordan Davis}, \bibinfo{person}{Lin Tan}, \bibinfo{person}{Petr Babkin}, {and} \bibinfo{person}{Sameena Shah}.} \bibinfo{year}{2023}\natexlab{}.
\newblock \showarticletitle{How effective are neural networks for fixing security vulnerabilities}. In \bibinfo{booktitle}{\emph{Proceedings of the 32nd ACM SIGSOFT International Symposium on Software Testing and Analysis}}. \bibinfo{pages}{1282--1294}.
\newblock


\bibitem[Xia et~al\mbox{.}(2023)]%
        {xia2023automated}
\bibfield{author}{\bibinfo{person}{Chunqiu~Steven Xia}, \bibinfo{person}{Yuxiang Wei}, {and} \bibinfo{person}{Lingming Zhang}.} \bibinfo{year}{2023}\natexlab{}.
\newblock \showarticletitle{Automated program repair in the era of large pre-trained language models}. In \bibinfo{booktitle}{\emph{2023 IEEE/ACM 45th International Conference on Software Engineering (ICSE)}}. IEEE, \bibinfo{pages}{1482--1494}.
\newblock


\bibitem[Xie et~al\mbox{.}(2023)]%
        {xie2023impact}
\bibfield{author}{\bibinfo{person}{Danning Xie}, \bibinfo{person}{Byungwoo Yoo}, \bibinfo{person}{Nan Jiang}, \bibinfo{person}{Mijung Kim}, \bibinfo{person}{Lin Tan}, \bibinfo{person}{Xiangyu Zhang}, {and} \bibinfo{person}{Judy~S Lee}.} \bibinfo{year}{2023}\natexlab{}.
\newblock \showarticletitle{Impact of large language models on generating software specifications}.
\newblock \bibinfo{journal}{\emph{arXiv preprint arXiv:2306.03324}} (\bibinfo{year}{2023}).
\newblock


\bibitem[Xie et~al\mbox{.}(2024)]%
        {xie2024resym}
\bibfield{author}{\bibinfo{person}{Danning Xie}, \bibinfo{person}{Zhuo Zhang}, \bibinfo{person}{Nan Jiang}, \bibinfo{person}{Xiangzhe Xu}, \bibinfo{person}{Lin Tan}, {and} \bibinfo{person}{Xiangyu Zhang}.} \bibinfo{year}{2024}\natexlab{}.
\newblock \showarticletitle{ReSym: Harnessing LLMs to Recover Variable and Data Structure Symbols from Stripped Binaries}.
\newblock  (\bibinfo{year}{2024}).
\newblock


\bibitem[Xu et~al\mbox{.}(2023)]%
        {xu2023improving}
\bibfield{author}{\bibinfo{person}{Xiangzhe Xu}, \bibinfo{person}{Shiwei Feng}, \bibinfo{person}{Yapeng Ye}, \bibinfo{person}{Guangyu Shen}, \bibinfo{person}{Zian Su}, \bibinfo{person}{Siyuan Cheng}, \bibinfo{person}{Guanhong Tao}, \bibinfo{person}{Qingkai Shi}, \bibinfo{person}{Zhuo Zhang}, {and} \bibinfo{person}{Xiangyu Zhang}.} \bibinfo{year}{2023}\natexlab{}.
\newblock \showarticletitle{Improving binary code similarity transformer models by semantics-driven instruction deemphasis}. In \bibinfo{booktitle}{\emph{Proceedings of the 32nd ACM SIGSOFT International Symposium on Software Testing and Analysis}}. \bibinfo{pages}{1106--1118}.
\newblock


\bibitem[Yang et~al\mbox{.}(2023)]%
        {yang2023asteria}
\bibfield{author}{\bibinfo{person}{Shouguo Yang}, \bibinfo{person}{Chaopeng Dong}, \bibinfo{person}{Yang Xiao}, \bibinfo{person}{Yiran Cheng}, \bibinfo{person}{Zhiqiang Shi}, \bibinfo{person}{Zhi Li}, {and} \bibinfo{person}{Limin Sun}.} \bibinfo{year}{2023}\natexlab{}.
\newblock \showarticletitle{Asteria-Pro: Enhancing Deep Learning-based Binary Code Similarity Detection by Incorporating Domain Knowledge}.
\newblock \bibinfo{journal}{\emph{ACM Transactions on Software Engineering and Methodology}} \bibinfo{volume}{33}, \bibinfo{number}{1} (\bibinfo{year}{2023}), \bibinfo{pages}{1--40}.
\newblock


\bibitem[You et~al\mbox{.}(2020)]%
        {you2020pmp}
\bibfield{author}{\bibinfo{person}{Wei You}, \bibinfo{person}{Zhuo Zhang}, \bibinfo{person}{Yonghwi Kwon}, \bibinfo{person}{Yousra Aafer}, \bibinfo{person}{Fei Peng}, \bibinfo{person}{Yu Shi}, \bibinfo{person}{Carson Harmon}, {and} \bibinfo{person}{Xiangyu Zhang}.} \bibinfo{year}{2020}\natexlab{}.
\newblock \showarticletitle{Pmp: Cost-effective forced execution with probabilistic memory pre-planning}. In \bibinfo{booktitle}{\emph{2020 IEEE Symposium on Security and Privacy (SP)}}. IEEE, \bibinfo{pages}{1121--1138}.
\newblock


\bibitem[Yu et~al\mbox{.}(2024)]%
        {yu2024llm}
\bibfield{author}{\bibinfo{person}{Jiahao Yu}, \bibinfo{person}{Xingwei Lin}, \bibinfo{person}{Zheng Yu}, {and} \bibinfo{person}{Xinyu Xing}.} \bibinfo{year}{2024}\natexlab{}.
\newblock \showarticletitle{$\{$LLM-Fuzzer$\}$: Scaling Assessment of Large Language Model Jailbreaks}. In \bibinfo{booktitle}{\emph{33rd USENIX Security Symposium (USENIX Security 24)}}. \bibinfo{pages}{4657--4674}.
\newblock


\bibitem[Zhang et~al\mbox{.}(2020)]%
        {zhang2020graph}
\bibfield{author}{\bibinfo{person}{Jiawei Zhang}, \bibinfo{person}{Haopeng Zhang}, \bibinfo{person}{Congying Xia}, {and} \bibinfo{person}{Li Sun}.} \bibinfo{year}{2020}\natexlab{}.
\newblock \showarticletitle{Graph-bert: Only attention is needed for learning graph representations}.
\newblock \bibinfo{journal}{\emph{arXiv preprint arXiv:2001.05140}} (\bibinfo{year}{2020}).
\newblock


\bibitem[Zhang et~al\mbox{.}(2021)]%
        {zhang2021osprey}
\bibfield{author}{\bibinfo{person}{Zhuo Zhang}, \bibinfo{person}{Yapeng Ye}, \bibinfo{person}{Wei You}, \bibinfo{person}{Guanhong Tao}, \bibinfo{person}{Wen-chuan Lee}, \bibinfo{person}{Yonghwi Kwon}, \bibinfo{person}{Yousra Aafer}, {and} \bibinfo{person}{Xiangyu Zhang}.} \bibinfo{year}{2021}\natexlab{}.
\newblock \showarticletitle{Osprey: Recovery of variable and data structure via probabilistic analysis for stripped binary}. In \bibinfo{booktitle}{\emph{2021 IEEE Symposium on Security and Privacy (SP)}}. IEEE, \bibinfo{pages}{813--832}.
\newblock


\bibitem[Zhao et~al\mbox{.}(2023)]%
        {zhao2023uvscan}
\bibfield{author}{\bibinfo{person}{Binbin Zhao}, \bibinfo{person}{Shouling Ji}, \bibinfo{person}{Xuhong Zhang}, \bibinfo{person}{Yuan Tian}, \bibinfo{person}{Qinying Wang}, \bibinfo{person}{Yuwen Pu}, \bibinfo{person}{Chenyang Lyu}, {and} \bibinfo{person}{Raheem Beyah}.} \bibinfo{year}{2023}\natexlab{}.
\newblock \showarticletitle{$\{$UVSCAN$\}$: Detecting $\{$Third-Party$\}$ Component Usage Violations in $\{$IoT$\}$ Firmware}. In \bibinfo{booktitle}{\emph{32nd USENIX Security Symposium (USENIX Security 23)}}. \bibinfo{pages}{3421--3438}.
\newblock


\bibitem[Zhou et~al\mbox{.}(2024)]%
        {zhou2024large}
\bibfield{author}{\bibinfo{person}{Xin Zhou}, \bibinfo{person}{Sicong Cao}, \bibinfo{person}{Xiaobing Sun}, {and} \bibinfo{person}{David Lo}.} \bibinfo{year}{2024}\natexlab{}.
\newblock \showarticletitle{Large Language Model for Vulnerability Detection and Repair: Literature Review and Roadmap}.
\newblock \bibinfo{journal}{\emph{arXiv preprint arXiv:2404.02525}} (\bibinfo{year}{2024}).
\newblock


\bibitem[Zhou et~al\mbox{.}(2019)]%
        {zhou2019devign}
\bibfield{author}{\bibinfo{person}{Yaqin Zhou}, \bibinfo{person}{Shangqing Liu}, \bibinfo{person}{Jingkai Siow}, \bibinfo{person}{Xiaoning Du}, {and} \bibinfo{person}{Yang Liu}.} \bibinfo{year}{2019}\natexlab{}.
\newblock \showarticletitle{Devign: Effective vulnerability identification by learning comprehensive program semantics via graph neural networks}.
\newblock \bibinfo{journal}{\emph{Advances in neural information processing systems}}  \bibinfo{volume}{32} (\bibinfo{year}{2019}).
\newblock


\end{thebibliography}

\end{document}